\documentclass[aps,prb,twocolumn,superscriptaddress,groupedaddress,showpacs]{revtex4}
\usepackage{graphicx,xcolor}
\usepackage{subfigure}
\usepackage{amsmath}
\usepackage{simplewick}
\usepackage{relsize}

\begin{document}

\title{Dual-fermion approach to the Anderson-Hubbard model}

\author{P.\ Haase}
\email{haasephysik@gmail.com}
\affiliation{Department of Physics, University of G\"ottingen, 37077 G\"ottingen, Germany}
\author{S.-X.\ Yang}
\email{yangphysics@gmail.com}
\affiliation{Department of Physics and Astronomy, Louisiana State University, Baton Rouge, Louisiana 70803, USA}
\affiliation{Center for Computation and Technology, Louisiana State University, Baton Rouge, Louisiana 70803, USA}
\author{T.\ Pruschke}
\altaffiliation{Deceased 12 January 2016.}
\affiliation{Department of Physics, University of G\"ottingen, 37077 G\"ottingen, Germany}
\author{J.\ Moreno}
\affiliation{Department of Physics and Astronomy, Louisiana State University, Baton Rouge, Louisiana 70803, USA}
\affiliation{Center for Computation and Technology, Louisiana State University, Baton Rouge, Louisiana 70803, USA}
\author{M.\ Jarrell}
\affiliation{Department of Physics and Astronomy, Louisiana State University, Baton Rouge, Louisiana 70803, USA}
\affiliation{Center for Computation and Technology, Louisiana State University, Baton Rouge, Louisiana 70803, USA}

\date{\today}

\begin{abstract}
We apply the recently developed dual fermion algorithm for disordered interacting systems to the Anderson-Hubbard model.
This algorithm is compared with dynamical cluster approximation calculations for a one-dimensional system to establish
the quality of the approximation in comparison with an established cluster method. We continue with a
three-dimensional (3d) system and look at the antiferromagnetic, Mott and Anderson localization transitions.
The dual fermion approach leads to quantitative as well as qualitative improvement of the dynamical mean-field results
and it allows one to calculate the hysteresis in the double occupancy in 3d taking into account nonlocal correlations.
\end{abstract}


\maketitle
\section{Introduction}
Electron-electron interactions have a strong impact on real materials, and the same holds true for disorder.
Both disorder and interaction can lead to localization, albeit the mechanism is quite different for both cases.
In correlated systems at half-filling, strong local Coulomb repulsion leads to localization as the double occupancy
of lattice sites becomes energetically too costly\cite{Mott}. In the strongly disordered systems, coherent 
backscattering leads to the localization of particles\cite{Anderson}.  Thus, it is not surprising that disordered 
interacting systems are an interesting topic to study\cite{Lee_RevModPhys,Belitz_RevModPhys}.

Both purely interacting and purely disordered systems pose challenges for theoretical treatment, especially in more 
than one and less than infinite dimensions. In one dimension, the Bethe ansatz\cite{Lieb1968} often allows for an 
analytic solution.  In infinite dimensions, dynamical mean-field theory\cite{DMFA1,DMFA2,DMFA3,DMFA_reviews1,
DMFA_reviews2} (DMFT) and the coherent potential approximation\cite{leath66,p_soven_67,d_taylor_67,shiba71} (CPA)
provide exact solutions for interacting and disordered systems, respectively. Jani\v{s} and Vollhardt\cite{Janis1992} 
extended DMFT to include both disorder and interaction.

The DMFT and CPA rely on a mapping of the lattice problem to an impurity problem that is solved self-consistently.
As a consequence of the local nature of the impurity problem, DMFT and CPA neglect nonlocal quantum
fluctuations altogether. Thus, these approaches are unreliable when it comes to systems with important 
nonlocal physics. To address this problem, a number of nonlocal extensions of DMFT have been devised.
These include the dynamical cluster approximation (DCA)\cite{Hettler98,Hettler00,m_jarrell_01a}, 
the traveling cluster approximation (TCA)\cite{mills78,kaplan80},
the molecular coherent potential approximation (MCPA)\cite{tsukada69,tsukada72,f_ducastelle_74}
and the cluster coherent potential approximation (CCPA)\cite{mookerjee73,kaplan76a,kaplan76b,kumar82,mookerjee87}.
Whereas DMFT uses a single impurity problem, the aforementioned methods use a finite cluster, 
which allows to take into account short range correlations.

A common roadblock of cluster methods for interacting systems is the solution of the interacting electron problem
on the cluster.  For weak interactions, perturbation theory can be applied, which is numerically feasible for 
relatively large system sizes.  For strong interactions, however, more elaborate cluster solvers like quantum Monte 
Carlo (QMC) \cite{m_jarrell_92a,m_jarrell_01c,Rubtsov1,Rubtsov2,Assaad07,Gull2008,Werner2006,GullMonteCarlo} are 
needed. The infamous sign problem limits the range of applicability of QMC to relatively small clusters and high 
temperatures. Even without the sign problem, it is difficult to solve large enough clusters with the precision 
needed for self-consistent methods like DCA.

A way out are diagrammatic extensions of DMFT\cite{Janis2001}, these include the dual fermion approach 
(DF)\cite{Rubtsov08}, the dynamical vertex approximation (D$\Gamma$A)\cite{Held2008}, and the multi-scale
many-body method\cite{c_slezak_06b}. Originally developed for interacting systems, Terletska et al.\cite{h_terletska_13} 
extended the dual fermion approach to treat disordered systems. We extended the approach to disordered interacting 
systems\cite{YangHaase} and applied it to the Anderson-Falicov-Kimball model.  The DF method relies on the introduction 
of new degrees of freedom which allow for an efficient perturbative treatment.  The perturbative expansion is done 
around an impurity problem which serves as a reference system. The hybridization function, and thus the somewhat 
optimal impurity problem is determined self-consistently, analogous to DMFT.

The DF method becomes particularly efficient in the context of disordered systems as the number of disorder realizations 
can be kept small. This becomes obvious for a discrete disorder distribution like binary disorder. 
There are only two realizations for an impurity problem but $2^{N_c}$ for a cluster with $N_c$ sites.
Even if only a random sample of configurations is picked, it will generally be much larger than two.
In our experience the cost for solving a small cluster is comparable to solving an impurity problem including the full 
impurity vertex, the reduced number of configurations makes DF more cost-efficient than DCA or other cluster methods.

The paper is organized as follows: 
In section \ref{sec:Formalism} we briefly introduce the dual fermion formalism for the Anderson-Hubbard model.
We explain the essentials of the dual fermion mapping and name the contributions to the dual potential. 
The discussion of the formalism is concluded by providing the formulas for the second-order and the fluctuation 
exchange (FLEX) approximations  for the dual self-energy.
In section \ref{sec:Results} we show results for the one- and three-dimensional Anderson-Hubbard model.
We start with the one-dimensional (1d) system, 
where our goal is not the comparison with exact results but rather a comparison with DCA to see how DF compares to 
established cluster methods. We continue with the three-dimensional (3d) system and explore the antiferromagnetic 
and Mott transitions. Finally, we calculate a phase diagram on the $UV$ plane, where $U$ parameterizes the Hubbard
interaction and $V$ the disorder.

\section{Formalism}
\label{sec:Formalism}
\subsection{Dual-fermion mapping}
We will apply the dual fermion formalism (DF) for disordered interacting systems to the Anderson-Hubbard model, 
which has the Hamiltonian
\begin{eqnarray}
H_{AH}=&-\mathlarger{\sum}\limits_{ij,\sigma}(t_{ij}+\mu\delta_{ij})(c^\dagger_{i \sigma} c^{\phantom\dagger}_{j \sigma}+h.c.)- \mathlarger{\sum}\limits_{i,\sigma} v_{i}n_{i\sigma}+\nonumber\\
&+ U\mathlarger{\sum}\limits_i (n_{i\uparrow}-\frac{1}{2})(n_{i\downarrow}-\frac{1}{2}).
\label{eq:AH}
\end{eqnarray}
Here, $t_{ij}$ is the hopping matrix element between sites $i$ and $j$, $\mu$ is the chemical potential, 
$c^{(\dagger)}_{i \sigma}$ destroys (creates) an electron of spin $\sigma$ at site $i$,
$n_{i \sigma}=c^\dagger_{i \sigma}c_{i \sigma }$ measures the occupation of site $i$ with an electron 
of spin $\sigma$ and $n_i=n_{i \uparrow}+n_{i \downarrow}$ measures the total occupancy at site $i$.
The two interaction terms in the Hamiltonian are the Hubbard term, which is parameterized by $U$
and the disorder term with a
random potential $v_i$ that is distributed according to a probability distribution $P(v_i)$.
In this paper we use a binary distribution
\begin{equation}
P_{\text{Bin}}(v_i)= \frac{1}{2}\Big[\delta\Big(v_i-\frac{V}{2}\Big)+\delta\Big(v_i+\frac{V}{2}\Big)\Big],
\end{equation}
and a box distribution 
\begin{equation}
P_{\text{Box}}(v_i)= \frac{1}{V}\Theta\Big(\frac{V}{2}-|v_i|\Big).
\end{equation}
$\Theta$ is the Heaviside function
\begin{equation}
\Theta(x)=\begin{cases}0 \text{ if } x<0\\
1 \text{ if } x\ge 0
\end{cases}
\end{equation}
and $V$ parameterizes the disorder strength.

The introduction of the dual degrees of freedom works very much the same as for the Anderson-Falicov-Kimball 
model as discussed in Yang et al.\cite{YangHaase}.  The difference is that here we have to deal with two types of 
charge carries, spin up and spin down electrons, that can interact with each other.  Unlike for the 
Anderson-Falicov-Kimball model, this interaction leads to an impurity vertex function that fully depends on 
three frequencies as the Hubbard interaction leads to dynamic electron-electron scattering.

Assuming spin symmetry, the ``Formalism'' section of Yang et al.\cite{YangHaase} remains valid for the 
Anderson-Hubbard model, except that the dual potential becomes spin dependent.  The dual fermion mapping is 
done in the usual way (c.\,f. Appendix \ref{sec:df-mapping}) and leads to the dual action
\begin{equation}
S_d[f,f^*]=-\sum_{\omega,k,\sigma}G^{-1}_{d0,\sigma}(\omega,k)f^*_{\omega,k,\sigma}f^{\phantom *}_{\omega,k,\sigma}+\sum_i V_{d,i}
\end{equation}
with the bare dual Green function
\begin{equation}
G_{d0,\sigma}(w,{\bf k})\equiv G_{lat,\sigma}(w,{\bf k})-G_{\sigma}(w).
\end{equation}
$G_{lat}$ is the lattice Green function and $G$ the impurity Green function.
The dual potential in the particle-particle channel reads
\begin{eqnarray}
 V^{pp}_{d,i} & = & \frac{1}{2}\sum_{w,w^{\prime},\sigma_1,\sigma_2}V^{p,0}_{\sigma_1,\sigma_2}(w,w^{\prime})\times\nonumber\\
& \times &f_{i,w,\sigma_1}^{*}f_{i,w^{\prime},\sigma_2}^{*}f^{\phantom *}_{i,w^{\prime},\sigma_2}f^{\phantom *}_{i,w,\sigma_1}
\nonumber \\
 & + & \frac{1}{4}\sum_{w,w^{\prime},\nu}\sum_{\sigma_1,\sigma_2,\sigma_3,\sigma_4}V^{p,1}_{\sigma_1,\sigma_2,\sigma_3,\sigma_4}(\nu)_{w,w^{\prime}}\times\nonumber\\
& \times & f_{i,w+\nu,\sigma_1}^{*}f_{i,-w,\sigma_2}^{*}f_{i,-w^{\prime},\sigma_3}f_{i,w^{\prime}+\nu,\sigma_4}.
\label{eq:dual_potential} 
\end{eqnarray}
$V^{p,0}$ is given by the purely disordered contributions to the full impurity vertex and 
$V^{p,1}$ is given by all other contributions to the full impurity vertex.
The prefactor $\frac{1}{2}$ is due to the lack of crossing-symmetry of $V^{p,0}$. 
The dual potential is discussed in more detail in part B of this section.

\begin{figure}[tbh]
\centerline{ \includegraphics[clip,scale=0.6]{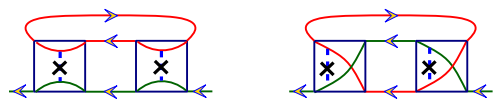}}
\caption{(Color online). Two second order diagrams for the self-energy in the particle-particle channel 
that contain closed loops (red lines). The diagram on the right contains the ``crossed'' disorder vertex.
}
\label{fig:closedLoops} 
\end{figure}

In the derivation of the formalism we use the replica trick as in Terleska et al.\cite{h_terletska_13}
It leads to the same restrictions for the diagrams as for the Anderson-Falicov-Kimball model\cite{YangHaase},
namely diagrams with closed Fermi loops that are only connected via disorder scattering are removed.
Two examples of what we call closed Fermi loops are given in Fig.~\ref{fig:closedLoops}.  A detailed 
discussion of how to obtain the final diagrams for the formalism from the replica trick is given in 
Appendix \ref{app:replim}.

\subsection{Dual potential}
The two-particle vertex has two very different contributions, one is crossing-symmetric, the other is crossing-asymmetric.
The crossing-asymmetric terms are given by the purely disordered contributions to the two-particle level. 
In terms of two-particle diagrams this means that the two single-particle Green function lines 
are connected by disorder-scattering only.
Since the dual potential has the full spin dependence as for the Hubbard model\cite{Rubtsov08}, we use an $SU(2)$ 
symmetric representation in terms of the density and magnetic channels 
\begin{equation}
V_{d/m^0}=V^{ph}_{\uparrow\uparrow;\uparrow\uparrow}\pm V^{ph}_{\uparrow\uparrow;\downarrow\downarrow}
\end{equation}
for the particle-hole channel, and the singlet and triplet channels
\begin{eqnarray}
V_{s/t^0}=V^{pp}_{\uparrow\downarrow;\uparrow\downarrow}\mp V^{pp}_{\uparrow\downarrow;\downarrow\uparrow}
\end{eqnarray}
for the particle-particle channel.

\begin{figure}[tbh]
\centerline{ \includegraphics[clip,scale=0.6]{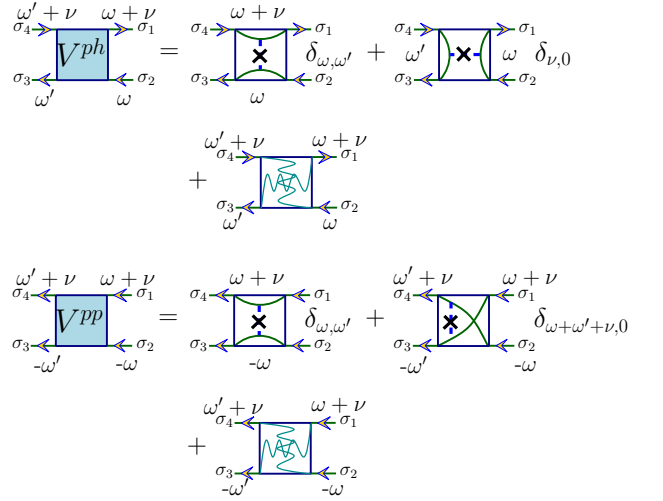}}
\caption{(Color online). Decomposition of the full vertex $V$ into two purely disordered contributions and all 
the rest for the particle-hole (top) and particle-particle channel (bottom). Along the green lines inside the boxes
spin and energy are conserved. The vertical (crossed) contribution for the particle-hole(particle) channel 
(second diagram in each case) is unphysical, but it is part of the vertex as defined in Appendix \ref{app:vertexFunctions}.
}
\label{fig:vertex_decomposed} 
\end{figure}

\begin{figure}[tbh]
\centerline{ \includegraphics[clip,scale=0.6]{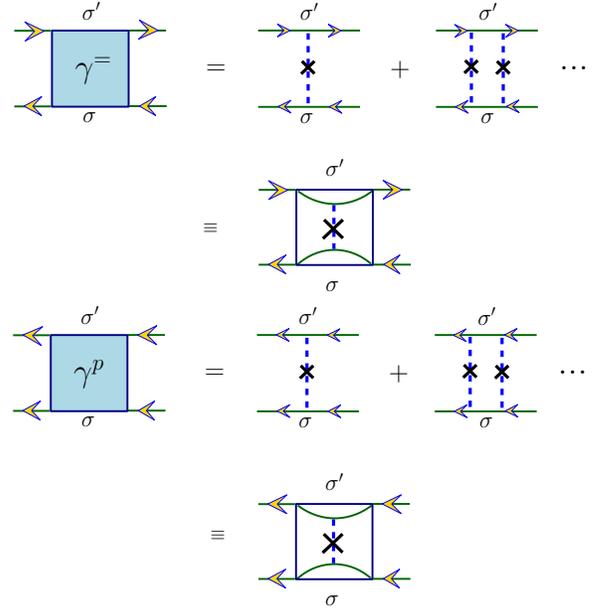}}
\caption{(Color online). Lower order contributions to the purely disordered vertex functions 
$\gamma^=$ and $\gamma^p$.}
\label{fig:gamma_parallel} 
\end{figure}

\begin{figure}[tbh]
\centerline{ \includegraphics[clip,scale=0.6]{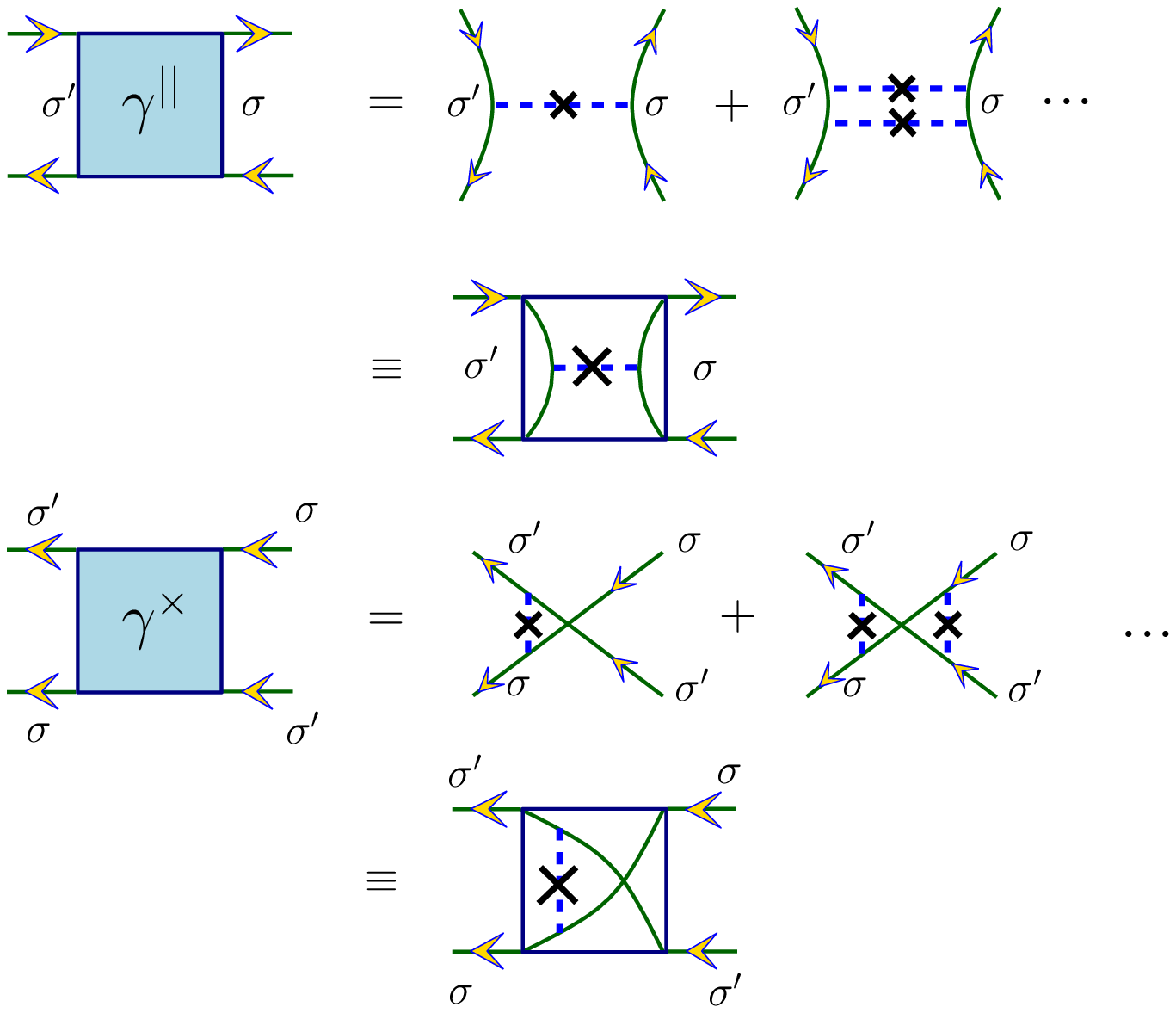}}
\caption{(Color online). Lower order contributions to the purely disordered vertex functions 
$\gamma^{||}$ and $\gamma^{\times}$.}
\label{fig:gamma_vertical} 
\end{figure}

\begin{figure}[tbh]
\centerline{ \includegraphics[clip,scale=0.6]{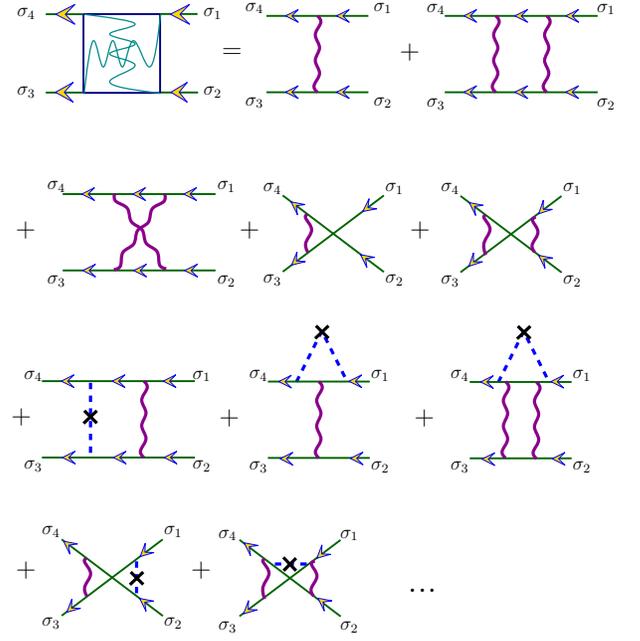}}
\caption{(Color online). Lower order contributions to the full vertex function as defined in App. \ref{app:vertexFunctions}.
The two diagrams at the bottom are would be unphysical without the Coulomb interaction lines. 
With the Coulomb lines the diagrams are physical which can be seen as follows:
before the disorder average, only Coulomb lines connect the two Green function lines. 
Additionally, the Green function lines include scattering from an arbitrary number of impurities,
in this case one scattering event for each Green function line.
Before the disorder average these scattering events are unconnected. After the disorder average,
the scattering events become connected and the above diagrams are created.
}
\label{fig:gamma_U_pp} 
\end{figure}

\begin{figure}[tbh]
\centerline{ \includegraphics[clip,scale=0.6]{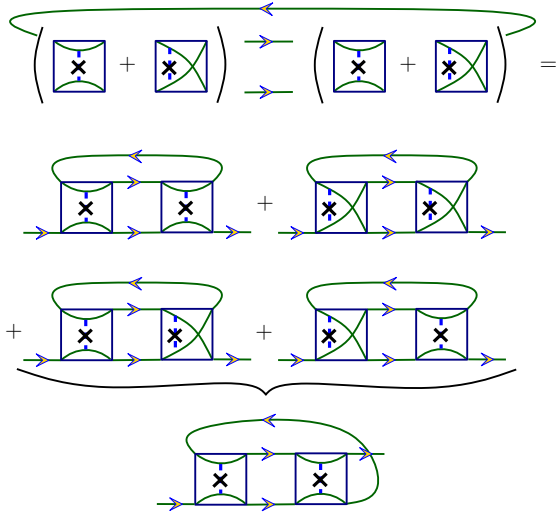}}
\caption{(Color online). Summing $\gamma^p$ and $\gamma^\times$ yields a crossing symmetric vertex function. 
This gives four different diagrams. The two diagrams in the middle are removed by the replica limit as there are closed loops. 
The two diagrams with $\gamma^p$ and $\gamma^\times$ are equivalent to the Fock-like diagram. In Eq. (\ref{eq:dual_potential}) 
the purely disordered part of the dual potential is crossing asymmetric and carries a factor $\frac{1}{2}$. 
If we replace it with the crossing symmetry disorder vertex a factor $\frac{1}{4}$ is needed to avoid double counting.}
\label{fig:crossing_symmetry} 
\end{figure}

The spin-dependent contributions $V^{ph(pp)}_{\sigma_1\sigma_2;\sigma_3\sigma_4}(\omega,\omega^\prime,\nu)$ to the 
dual potential are calculated from the disorder-averaged two-particle Green function, which is shown in 
Appendix \ref{app:vertexFunctions}.  These quantities are illustrated in Fig. \ref{fig:vertex_decomposed} and some 
lower order diagrams are shown in Figs. \ref{fig:gamma_parallel} to \ref{fig:gamma_U_pp}. With the measurement 
formulas (\ref{eq:v_ph_appendix}) and (\ref{eq:v_pp_appendix}) there are three different contributions that we have 
to distinguish. The purely disordered vertical (cross) channel is unphysical, but we find it convenient to keep it,
as it allows to restrict oneself to Hartree-like diagrams, 
which is illustrated in Fig. \ref{fig:crossing_symmetry}.
This is based on the fact
that for a crossing-symmetric interaction Hartree- and Fock-like diagrams are equivalent. 
The purely disordered vertex function becomes crossing-symmetric if one adds the
vertical (cross) channel to the horizontal channel. Let us stress that it is not recommended
to combine them into one symbol (numerically and diagrammatically) as these contributions and the
resulting diagrams behave very differently in the replica limit.

The purely disordered contributions to the full vertex will generally lead to unphysical diagrams 
and it is show in Section (\ref{sec:dfSelfEnergy}) how to remove them.
To this end  we introduce 
\begin{eqnarray}
V^0_{d}&=&V^{ph,0}_{\uparrow\uparrow;\uparrow\uparrow}+ V^{ph,0}_{\uparrow\uparrow;\downarrow\downarrow}=\gamma^=+2\gamma^{||}\label{eq:V_0_d}\label{eq:Vd0}\\
V^0_{m^0}&=&V^{ph,0}_{\uparrow\uparrow;\uparrow\uparrow}- V^{ph,0}_{\uparrow\uparrow;\downarrow\downarrow}=\gamma^=\label{eq:Vm0}
\end{eqnarray}
and
\begin{eqnarray}
V^0_{s}&=&V^{pp,0}_{\uparrow\downarrow;\uparrow\downarrow}- V^{pp,0}_{\uparrow\downarrow;\downarrow\uparrow}=\gamma^p-\gamma^\times\\
V^0_{t^0}&=&V^{pp,0}_{\uparrow\downarrow;\uparrow\downarrow}+ V^{pp,0}_{\uparrow\downarrow;\downarrow\uparrow}=\gamma^p+\gamma^\times\label{eq:V_0_t}
\end{eqnarray}
for the purely disordered contributions, where $\gamma^c$ is the disorder vertex functions for the channel $c$.
The different $\gamma^c$ are illustrated in terms of lower order diagrams in Figs. \ref{fig:gamma_parallel} and \ref{fig:gamma_vertical}.
All $\gamma^c$ are independent of the spin configuration as is indicated in Eqs. \ref{eq:V_0_d} to \ref{eq:V_0_t}.
Note that $\gamma^{||}$ appears only in $V^0_d$, but with a factor of $2$. 
Taking together the definition of $V_d^0$ and Figs. \ref{fig:vertex_decomposed}, \ref{fig:gamma_parallel} and \ref{fig:gamma_vertical},
it becomes clear that this is because $V^{ph(,0)}_{\uparrow\uparrow;\uparrow\uparrow}$ has a horizontal and a vertical contribution,
whereas $V^{ph(,0)}_{\uparrow\uparrow;\downarrow\downarrow}$ only has a vertical one.

The purely disordered contributions depend only on two frequencies, 
either two fermionic frequencies or one fermionic and the other bosonic.
For the first case, $\gamma^=(\omega,\omega^\prime)$ is obtained according to
\begin{eqnarray}
\gamma^=(\omega,\omega^\prime)&
=&\frac{1}{T}\frac{\{ g_{\sigma}(\omega)g_{\sigma^\prime}(\omega^\prime)\}-G_\sigma(\omega)G_{\sigma^\prime}(\omega^\prime)}
{G_{\sigma}(\omega)G_{\sigma^\prime}(\omega^\prime)G_{\sigma}(\omega)G_{\sigma^\prime}(\omega^\prime)}.
\label{eq:gamma_parallel}
\end{eqnarray}

Alternatively, we can calculate $\gamma^=$ using one fermionic and one bosonic frequency according to
\begin{equation}
\gamma^=(\nu)_\omega= \frac{1}{T}
\frac{\{ g_{\sigma}(\omega)g_{\sigma^\prime}(\omega+\nu)\}-G_\sigma(\omega)G_{\sigma^\prime}(\omega+\nu)}
{G_{\sigma}(\omega)G_{\sigma^\prime}(\omega+\nu)G_{\sigma}(\omega)G_{\sigma^\prime}(\omega+\nu)}.
\end{equation}
It is convenient to have both representations at ones disposal.
The disorder two-particle Green function for the particle-particle channel can be calculated according to
\begin{equation}
\gamma^p(\nu)_\omega=\frac{1}{T}\frac{\{g_{\sigma}(-\omega) g_{\sigma^\prime}(\omega+\nu)\}-G_{\sigma}(-\omega) G_{\sigma^\prime}(\omega+\nu)}
{G_{\sigma}(-\omega) G_{\sigma^\prime}(\omega+\nu)G_{\sigma}(-\omega) G_{\sigma^\prime}(\omega+\nu)} .
\label{eq:gamma_p}
\end{equation}
On the r.h.s. of Eqs. \ref{eq:gamma_parallel} to \ref{eq:gamma_p} the spin labels $\sigma$ and $\sigma^\prime$ appear.
As noted above, the $\gamma$ are independent of the spin, but in a Monte Carlo calculation 
the spin still has to be considered. In practice, we average over all possible spin configurations
to improve the Monte Carlo estimate.

The three frequency representations of the crossing-asymmetric vertex functions are obtained according to
\begin{eqnarray}
V^0_d(\nu)_{\omega,\omega^\prime}&=& \gamma^=(\nu)_\omega \delta_{\omega,\omega^\prime}+2\gamma^{||}(\nu)_{\omega,\omega^\prime} \\
V^0_m(\nu)_{\omega,\omega^\prime}&=& \gamma^=(\nu)_\omega  \delta_{\omega,\omega^\prime} \\
V^0_s(\nu)_{\omega,\omega^\prime}&=& \gamma^p(\nu)_\omega \delta_{\omega,\omega^\prime}-\gamma^\times(\nu)_{\omega,\omega^\prime}  \\
V^0_t(\nu)_{\omega,\omega^\prime}&=& \gamma^p(\nu)_\omega \delta_{\omega,\omega^\prime}+\gamma^\times(\nu)_{\omega,\omega^\prime}  ,
\end{eqnarray}
where
\begin{eqnarray}
\gamma^{||}(\nu)_{\omega,\omega^\prime}&=&-\gamma^=(\omega-\omega^\prime)_{\omega^\prime}\delta_{\nu,0}\\
\gamma^\times(\nu)_\omega&=&-\gamma^p(\nu)_{-\omega-\nu}\delta_{\omega+\omega^\prime+\nu,0}.
\end{eqnarray}
This follows from exchanging two corners of the box for the vertex function to obtain $\gamma^{||}(\gamma^\times)$ from $\gamma^=(\gamma^p)$.

\subsection{Dual self-energy}
\label{sec:dfSelfEnergy}
The dual self-energy is obtained using perturbation theory and can 
in general be calculated according to
\begin{eqnarray}
\Sigma(\omega,k)  = & - &\frac{T}{N_{c}}\sum_{v;q}G(w+v,k+q)\Phi(v,q)_{w,w}\nonumber \\
 & + & \frac{T}{N_{c}}\sum_{v;q}G(-w+v,-k+q)\Phi^{p}(v,q)_{w,w}\nonumber \\
 & + & \frac{T}{N_{c}}\sum_{q}G(w,k+q)\Phi^{0}(w,w;q)\nonumber \\
 & + & \frac{T}{N_{c}}\sum_{q}G(w,-k+q)\Phi^{0,p}(w,w;q),
 \label{eq:sigma}
\end{eqnarray}
where $\Phi^{(p)}$ is the effective interaction for the particle-hole (particle) channel with the purely disordered contributions removed.
$\Phi^{0(,p)}$ contains the purely disordered contributions from the particle-hole (particle) channel.
The exact form of $\Phi^{(p)}$ and $\Phi^{0(,p)}$ depends on the approximation that is used to calculate the self-energy.

In Eq. \ref{eq:sigma} one has to avoid double counting. 
In first and second order the particle-hole and particle-particle diagrams are equivalent 
and hence only one channel must be used, 
e.\,g. this implies that the second order contribution of either the particle-hole or particle-particle channel has to be removed explicitly 
from the vertex ladder $\Phi$ for the fluctuation exchange approximation (FLEX).
The self-consistency condition removes all first order contributions, 
thus we will not consider them here.

To second order, the effective interaction for the particle-hole channel reads
\begin{equation}
\Phi=\frac{1}{4}[V_{d}\bar\chi^{ph}_{0}V_{d}+3V_{m}\bar\chi^{ph}_{0}V_{m}]-\frac{1}{4}[V_{d}^{0}\bar\chi^{ph}_{0}V_{d}^{0}+3V_{m}^{0}\bar\chi^{ph}_{0}V_{m}^{0}],
\label{eq:2ndPH}
\end{equation}
which has been calculated from the diagrams in Fig. \ref{fig:2ndOrderPH}. This is discussed in more detail 
in Appendix \ref{app:dualSelfEnergy}. Matrix multiplication is implied.  The corresponding disorder contribution is
\begin{equation}
\Phi^{0}(w,w;q)=\gamma^=(w,w)\bar{\chi}_{0}^{ph}(\nu=0;q)_\omega\gamma^=(w,w)
\end{equation}
and 
\begin{equation}
\bar{\chi}_{0}^{ph}(\nu,q)_\omega=\frac{T}{N}\sum_k G^d(\omega+\nu,k+q)G^d(\omega,k).
\end{equation}

\begin{figure}[tbh]
\centerline{ \includegraphics[clip,scale=0.6]{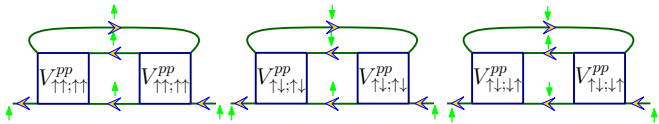}}
\caption{(Color online). Second order diagrams for the particle-particle channel.
For the first diagram, the two internal Green function lines are equivalent, therefore the diagram
comes with a symmetry factor $\frac{1}{2}$.}
\label{fig:2ndOrderPP} 
\end{figure}

Alternatively, the second-order self-energy can be calculated from the particle-particle channel. Fig.~\ref{fig:2ndOrderPP} 
shows the corresponding diagrams.  The effective interaction for the interacting disordered part reads
\begin{equation}
\Phi^{pp}=\frac{1}{2}[V_{s}\bar\chi^{p}_{0}V_{s}+3V_{t}\bar\chi^{p}_{0}V_{t}]-\frac{1}{2}[V_{s}^{0}\bar\chi^{p}_{0}V_{s}^{0}+3V_{t}^{0}\bar\chi^{p}_{0}V_{t}^{0}]
\end{equation}
and 
\begin{equation}
\Phi^{0,p}(w,w;q)=\gamma^p(w,w)\bar{\chi}_{0}^{p}(\nu=0,q)_\omega\gamma^p(w,w)
\end{equation}
for the purely disordered part with
\begin{equation}
\bar{\chi}_{0}^{p}(\nu,q)_\omega=-\frac{T}{2N}\sum_k G^d(\omega+\nu,k+q)G^d(-\omega,-k).
\end{equation}

It is also possible to sum ladder diagrams up to infinite order. This is done using FLEX for the dual degrees of freedom.
To this end, we need the vertex ladders for the particle-hole channel
\begin{equation}
F_{d/m}=\frac{V_{d/m}}{1-V_{d/m}\bar\chi^{ph}_{0}}
\end{equation}
and for the particle-particle channel
\begin{equation}
F_{s/t}=\frac{V_{s/t}}{1-V_{s/t}\bar\chi^{p}_{0}}.
\end{equation}

For the particle-hole channel we obtain
\begin{equation}
\Phi^{ph*}=\frac{1}{2}[V_{d}\bar\chi^{ph}_{0}(F_{d}-V_{d})+3V_{m}\bar\chi^{ph}_{0}(F_{m}-V_{m})].
\end{equation}
In the above, the second-order contribution has been removed. We can added it back and we obtain the right prefactor 
(cf. Eq.~\ref{eq:2ndPH}) by using
\begin{equation}
\Phi^{ph}=\frac{1}{4}[V_{d}\bar\chi^{ph}_{0}(2F_{d}-V_{d})+3V_{m}\bar\chi^{ph}_{0}(2F_{m}-V_{m})].
\end{equation}
Subtracting the purely disordered contributions we obtain
\begin{eqnarray}
\Phi&=&\frac{1}{4}[V_{d}\bar\chi^{ph}_{0}(2F_{d}-V_{d})+3V_{m}\bar\chi^{ph}_{0}(2F_{m}-V_{m})]\nonumber\\
&-&\frac{1}{4}[V_{d}^{0}\bar\chi^{ph}_{0}(2F_{d}^{0}-V_{d}^{0})+3V_{m}^{0}\bar\chi^{ph}_{0}(2F_{m}^{0}-V_{m}^{0})].
\end{eqnarray}
The physical disorder contributions for the particle-hole channel are given by
\begin{equation}
\Phi^0(w,w;q)=\gamma^=(1-\gamma^=\bar{\chi}_{0}^{ph})^{-2}-\gamma^=(1+\gamma^=\bar{\chi}_{0}^{ph}).
\end{equation}

In FLEX, both the particle-hole and particle-particle channel are used. 
The interacting and  disordered contributions are calculated according to
\begin{eqnarray}
\Phi^{pp}&=&\frac{1}{2}[V_{s}\bar\chi^{p}_{0}(F_{s}-V_{s})+3V_{t}\bar\chi^{p}_{0}(F_{t}-V_{t})]\nonumber \\
&-&\frac{1}{2}[V_{s}^{0}\bar\chi^{p}_{0}(F_{s}^{0}-V_{s}^{0})+3V_{t}^{0}\bar\chi^{p}_{0}(F_{t}^{0}-V_{t}^{0})]
\end{eqnarray}
for the particle-particle channel.  The corresponding disorder contribution is
\begin{equation}
\Phi^{0,p}(w,w;q)=\gamma^p(1-\gamma^p\bar{\chi}_{0}^{pp})^{-1}-\gamma^p(1+\gamma^p\bar{\chi}_{0}^{pp}).
\end{equation}

\begin{figure}[tbh]
\centerline{ \includegraphics[clip,scale=0.6]{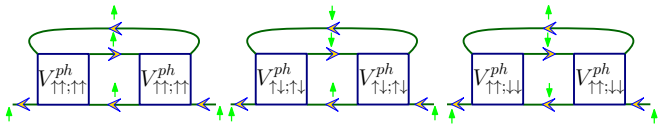}}
\caption{(Color online). Second order diagrams for the particle-hole channel. The second and third
diagram are topologically equivalent, therefore a symmetry factor $\frac{1}{2}$ is associated with these diagrams.}
\label{fig:2ndOrderPH} 
\end{figure}

\section{Results}
\label{sec:Results}
In this section we present results for the Anderson-Hubbard model.  We start with the one-dimensional (1d) 
system where we 
compare DF with DCA to see how the dual fermions compare to an established cluster method.  Next, we 
take a look at the three-dimensional (3d) system. First, we study the influence of disorder on the antiferromagnetic 
transition and how nonlocal correlations change the result.
Second, we take a look at the Mott transition. To this end, we take a look at the hysteresis of the double 
occupancy $D$ as a function of the Hubbard coupling $U$ and temperature $T$ and have a look at the effect of disorder.
Third, we calculate a phase diagram on the Hubbard and disorder strength ($UV$) plane.

All results are at half-filling and, if not otherwise stated, for binary disorder.

\subsection{Relative corrections for the 1d system}
For one dimension it is possible to obtain DCA results for disordered interacting systems at reasonable computational cost. 
The DCA results serve as a benchmark for the dual fermion results. We take a look at the relative correction
to the local Green function $G_{\text{loc}}$
\begin{equation}
\sigma(G_{\text{loc}})=\frac{\text{Im} G^{\text{\text{nloc}}}_{\text{loc}}(i\pi T)-\text{Im} G^{\text{DMFT}}_{\text{loc}}(i\pi T)}{|\text{Im} G^{\text{DMFT}}_{\text{loc}}(i\pi T)|},
\end{equation}
where $nloc$ refers to the result from the non-local method, either dual fermion or DCA. 
We use a self-consistent 2nd-order approximation as well as a FLEX approximation for the solution of the dual fermion problem.
In the following, the former will be referred to as DF-2nd, while the latter as DF-FLEX.

\begin{figure}[tbh]
\centerline{ \includegraphics[clip,scale=0.35]{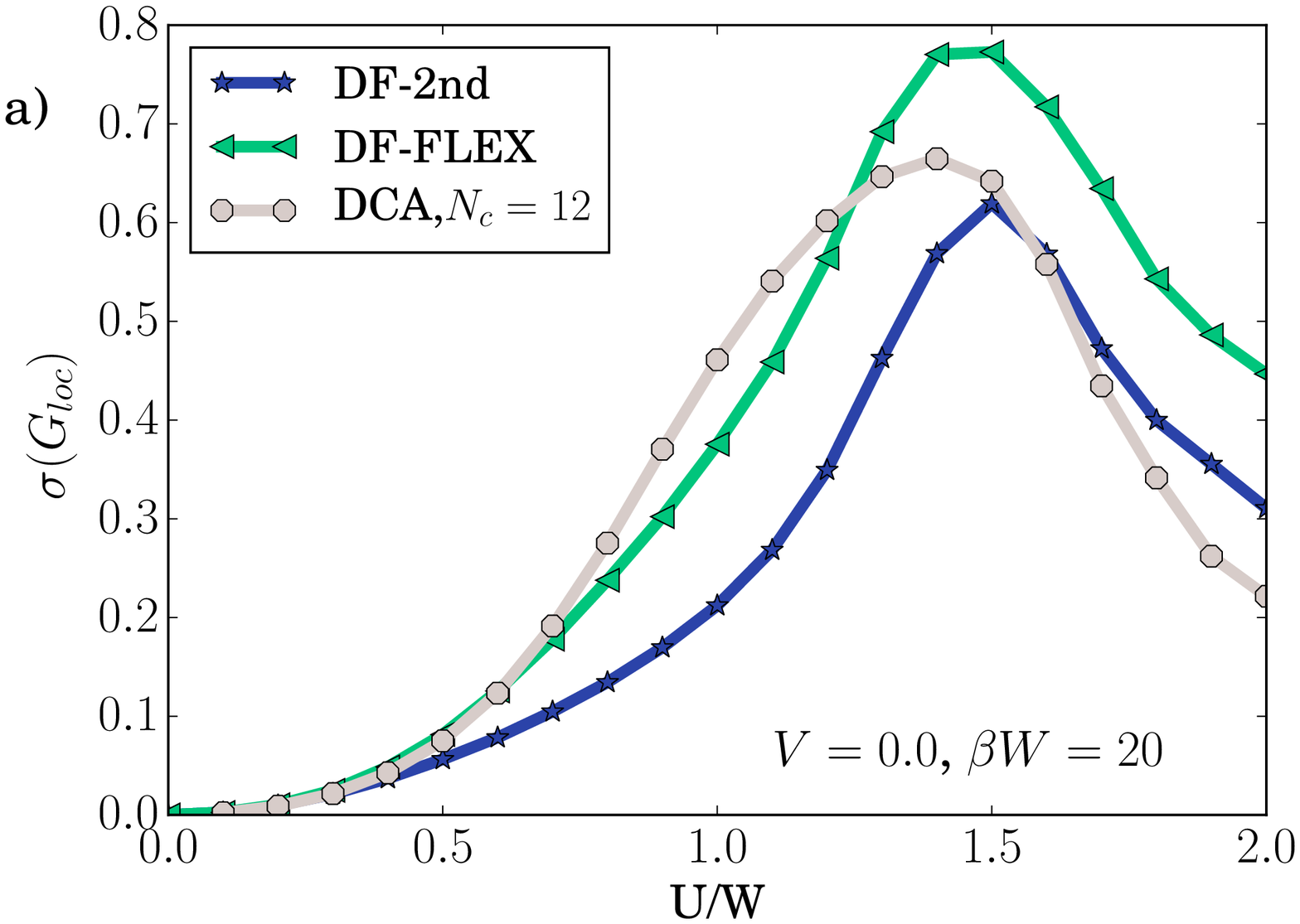}}

\centerline{ \includegraphics[clip,scale=0.35]{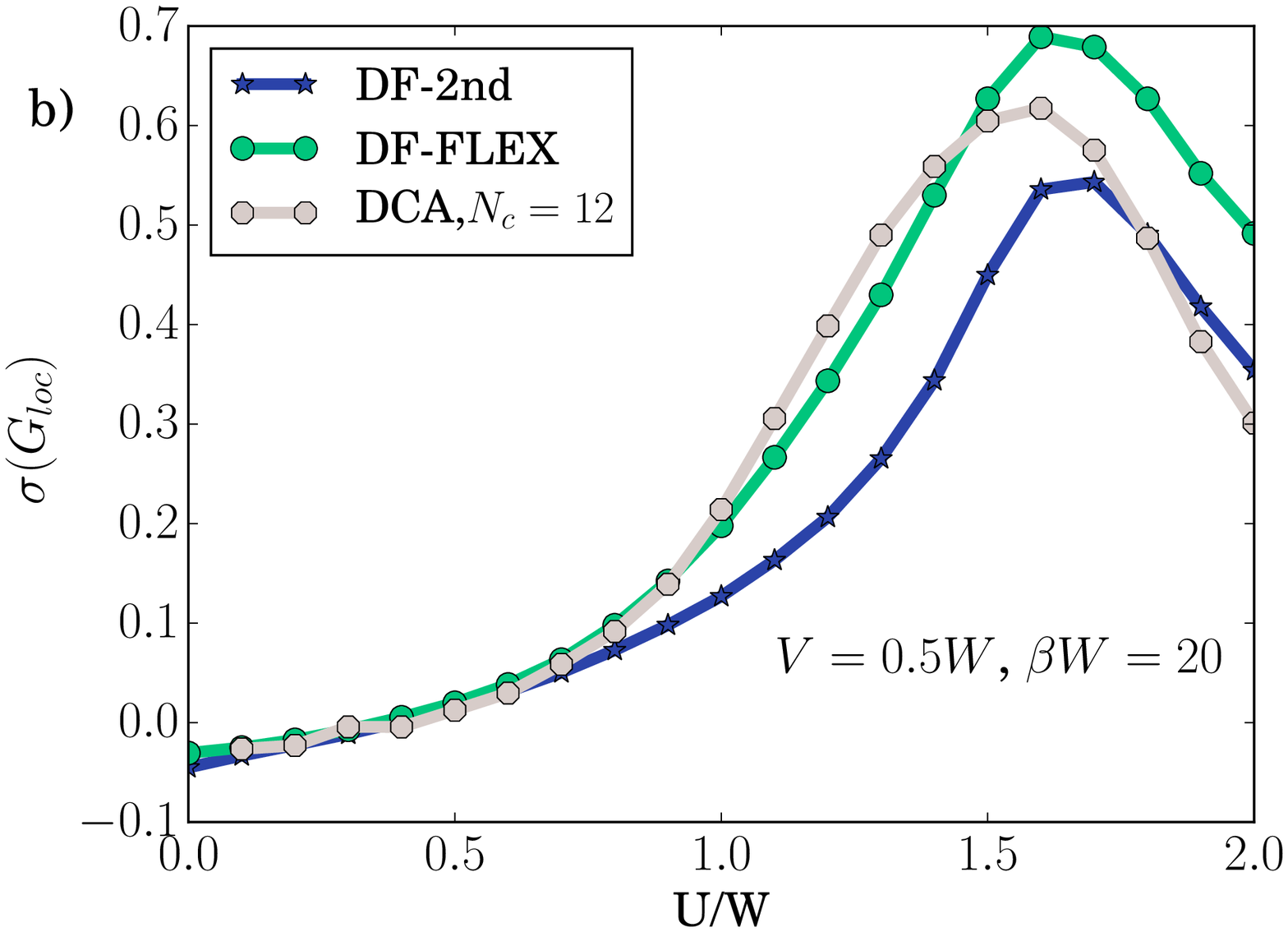}}

\centerline{ \includegraphics[clip,scale=0.35]{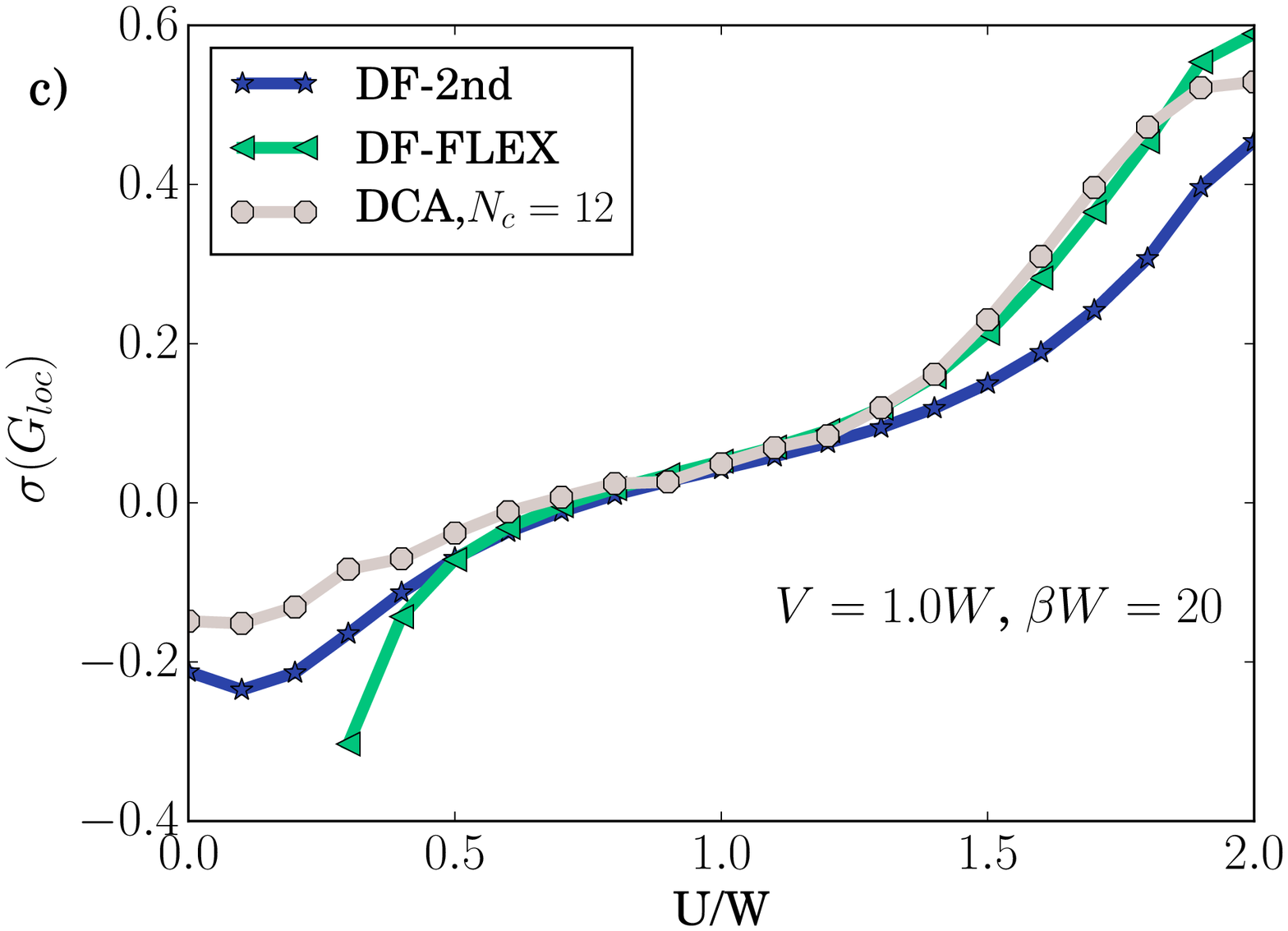}}
\caption{(Color online). Relative correction from the dual fermion approach
to the local Green function at the lowest Matsubara frequency ($iw=i\pi T$)
for various parameters of the 1d lattice. The corrections
are minimized for both weak- and large-U limits and maximized for values of U around
the band-with. The peak position shifts to larger U with
increasing disorder strength. This behavior is consistent with DCA results.}

\label{fig:gl-diff} 
\end{figure}

Results are shown in Fig.~\ref{fig:gl-diff}. We observe that the dual fermion results qualitatively agree 
with the DCA results for a 12-site cluster, which is a converged DCA solution. 
For the clean system, i.\,e. $V=0$, the maximum corrections are around $1.5\,W$, where $W$ is the bandwidth. 
The maximum corrections appear around the Mott transition, because the DF method gives a smaller critical $U$ 
than DMFT.  With increasing disorder strength the maximum corrections are moved to larger values of $U$ and the 
magnitude of the corrections are reduced. This is true for the DCA, DF-2nd and DF-FLEX. 
For $V=W$ and small $U$ DF-FLEX becomes unreliable and does not converge for $U\rightarrow 0$.
We conclude that the DF-FLEX agrees very well with the DCA below the $U$ of the maximum corrections (if applicable). 
For larger values of $U$ the DF-2nd method shows better agreement with the DCA.

\begin{figure}[tbh]
\centerline{ \includegraphics[clip,scale=0.35]{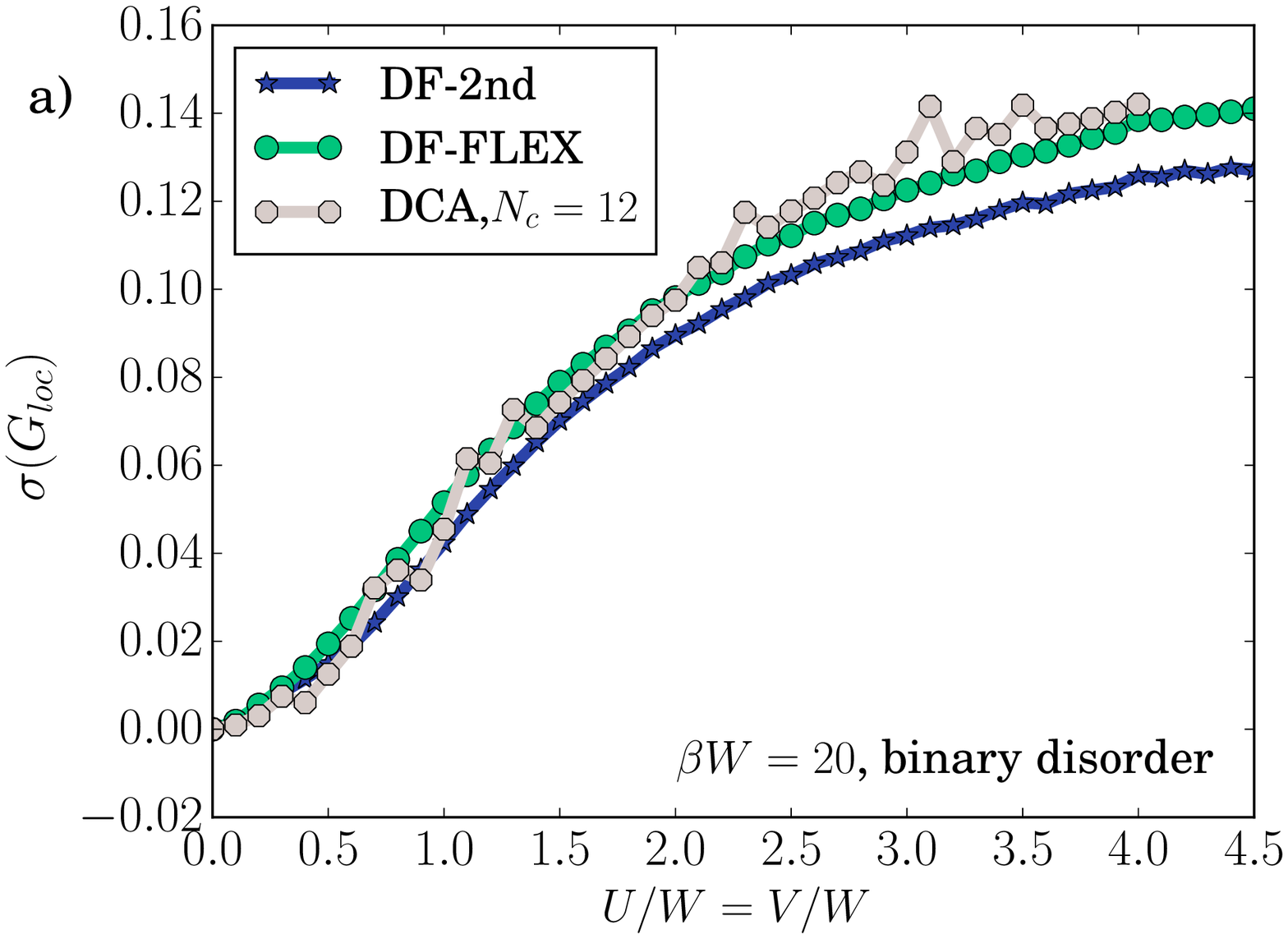}}

\centerline{ \includegraphics[clip,scale=0.35]{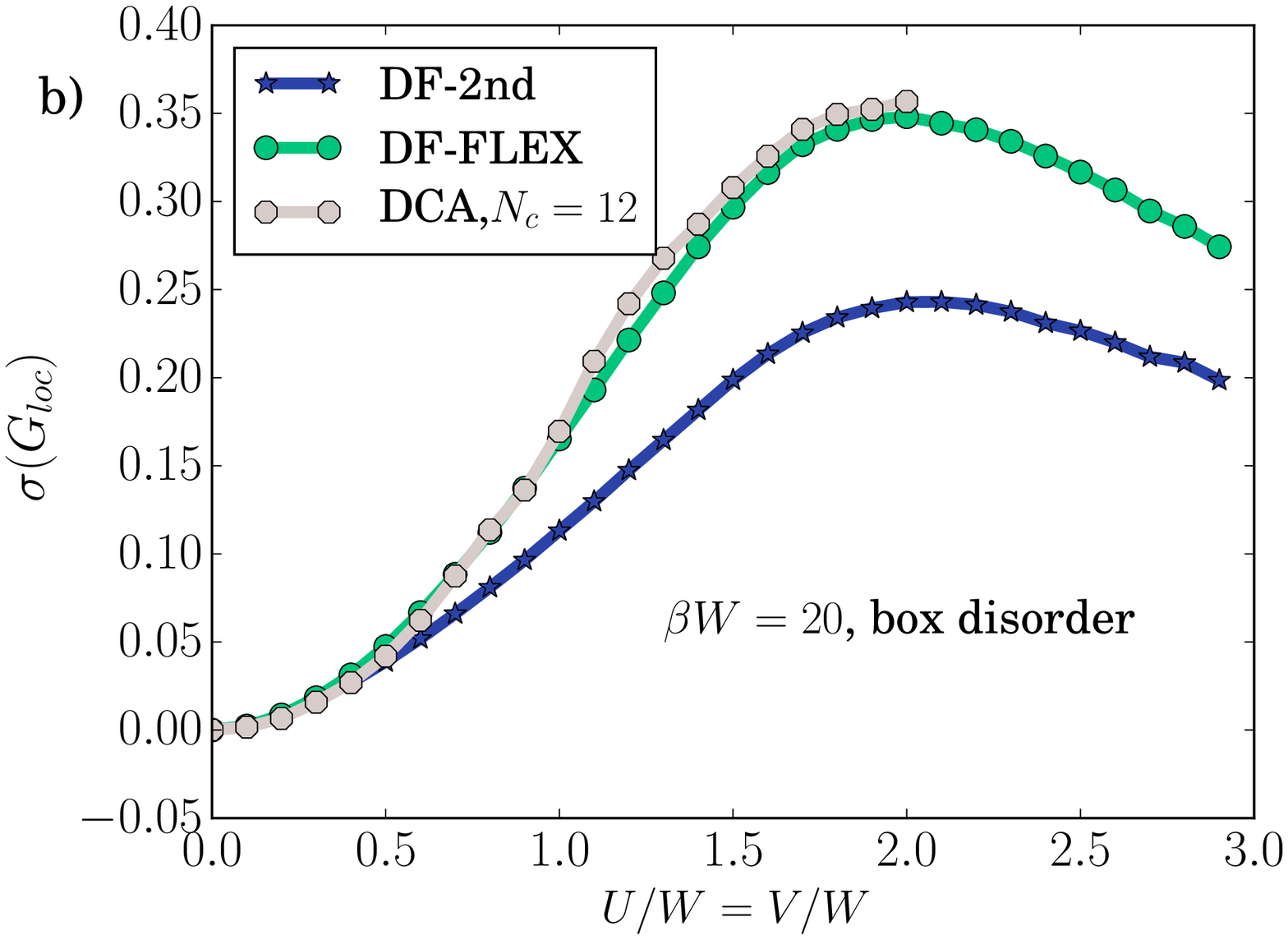}}

\caption{(Color online). Relative correction from the dual fermion approach
to the local Green function at the lowest Matsubara frequency ($iw=i\pi T$) for $U=V$.
In this case, the nonlocal corrections are strongly reduced by the disorder. 
In the first figure the dual fermion corrections agree quite well with the DCA correction. 
This is true for DF-2nd and DF-FLEX. For the second figure, the DF-2nd corrections
qualitatively reproduce DCA results, and the matching of DF-FLEX to DCA results is nearly perfect. 
}

\label{fig:gl-diffUV} 
\end{figure}

In Fig.~\ref{fig:gl-diffUV} we take a look at the special case $U=V$. 
We find for both binary and box disorder remarkable agreement between the DF-FLEX and DCA. 
DF-2nd agrees qualitatively, but there is a substantial quantitative deviation, especially for box disorder.
For both types of disorder, the sign problem limits the parameter range for which we can obtain DCA results. 
Also, the DCA results for binary disorder are quite noisy.
These results show the power of the DF method. When cluster methods become inefficient or not applicable at all, 
the DF method can often still be applied.

\subsection{Antiferromagnetic transition in the 3d system}
The 3d Hubbard model has an antiferromagnetic phase at finite temperatures. We investigate how the antiferromagnetic 
region changes when disorder is introduced and what happens if nonlocal correlations are taken into account.

\begin{figure}[tbh]
\centerline{ \includegraphics[clip,scale=0.35]{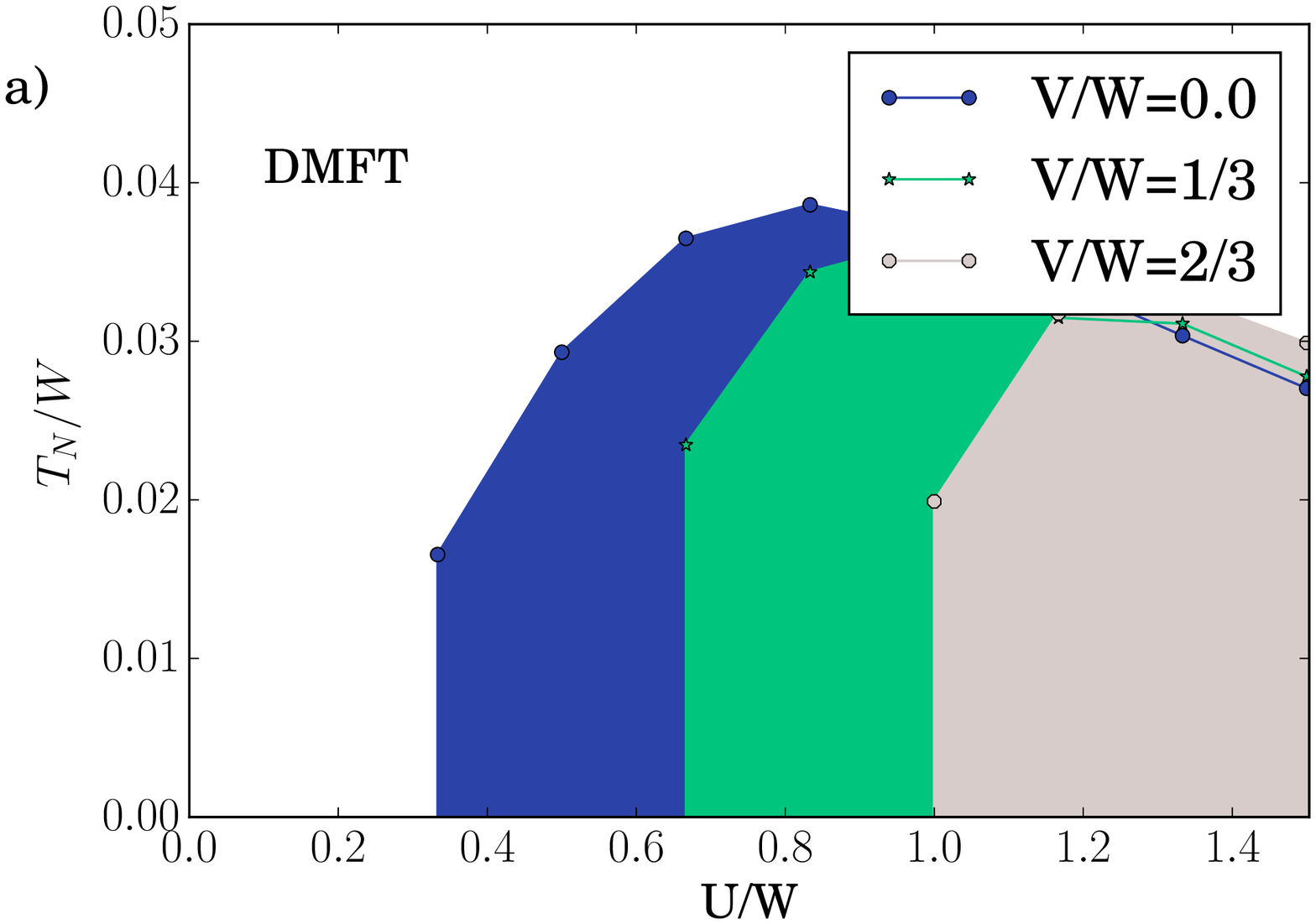}}

\centerline{ \includegraphics[clip,scale=0.35]{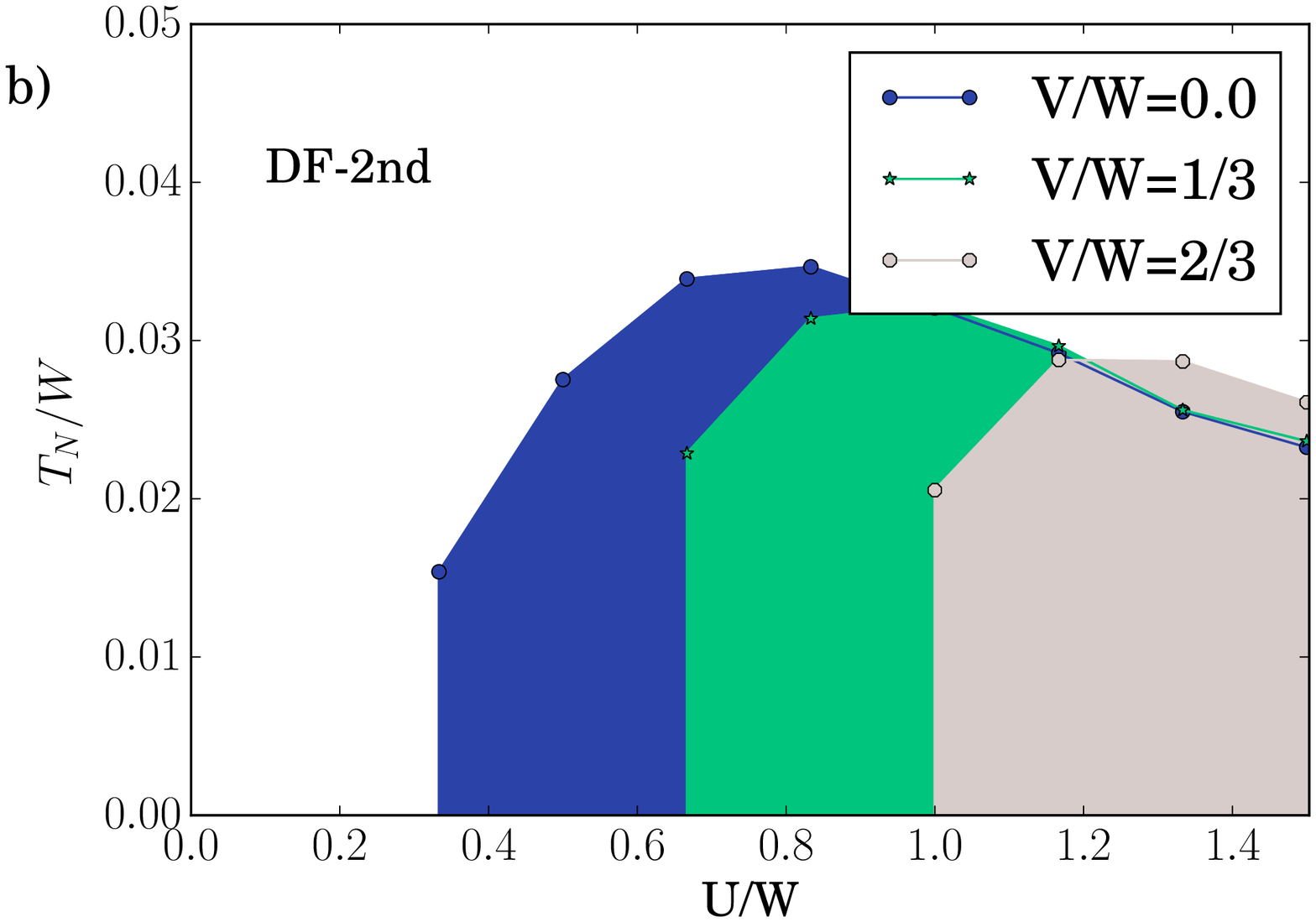}}

\centerline{ \includegraphics[clip,scale=0.35]{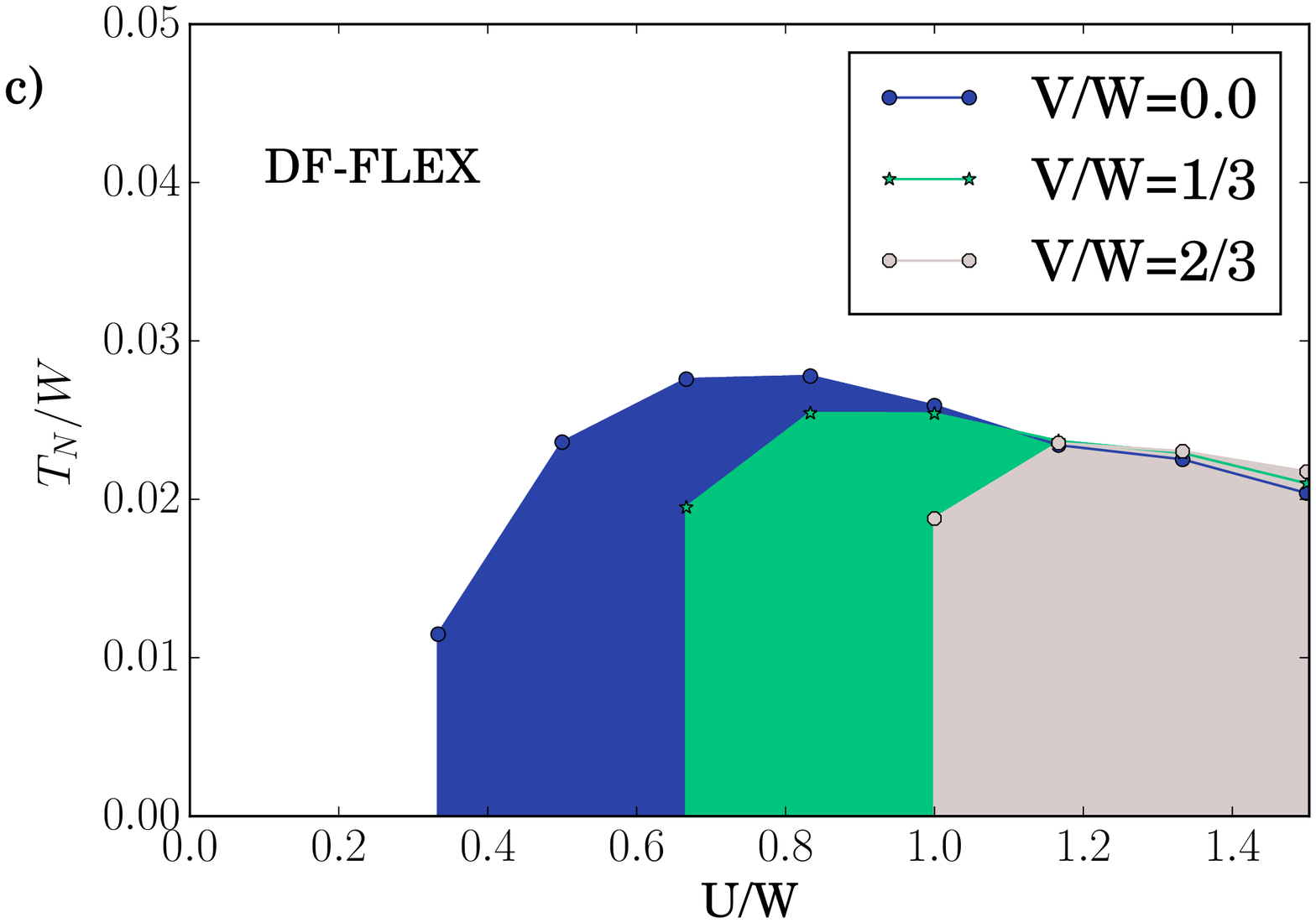}}
\caption{(Color online). The phase diagram of the 3d Anderson-Hubbard model on the $UT$-plane for various values of $V$
calculated with the DMFT+CPA, and DF-2nd and DF-FLEX dual fermion approaches.  For small values of $U$, the 
antiferromagnetic phase is suppressed by disorder.  For large values of $U$ the disorder increases $T_{N}$.  
The effect of dual fermions is to decrease the transition temperature. For the DF-FLEX approximation this effect is 
more pronounced.}

\label{fig:tcBin} 
\end{figure}

The antiferromagnetic phase transition is characterized by a divergence of the antiferromagnetic susceptibility. This 
is equivalent to a leading eigenvalue (LEV) for the Bethe-Salpeter equation that is equal to one. Therefore, we use the 
LEV to determine the antiferromagnetic phase boundary. Results are shown in Fig.~\ref{fig:tcBin}.  DMFT, DF-2nd and 
DF-FLEX give the same general solution. Disorder suppresses antiferromagnetism for small $U$. For large $U$ weak 
disorder enhances antiferromagnetism.
This agrees with the findings of Ulmke et al.\cite{Ulmke95} for the infinite dimensional Anderson-Hubbard model on the Bethe lattice. 

The reduction of the antiferromagnetic transition temperature for small $U$ agrees with the general expectation that disorder 
obstructs long-range order.  Ulmke et. al. \cite{Ulmke95} give an explanation for the increase of $T_{N}$ with disorder for 
large values of $U$. The argument is that virtual hopping processes between sites $A$ and $B$ leads to an energy gain 
$J_1=-t^2/[U-(\epsilon_A-\epsilon_B)]$ if $B$ is occupied by an electron of opposite spin and an energy gain 
$J_2=-t^2/[U+(\epsilon_A-\epsilon_B)]$ for hopping from $B$ to $A$. 
The relative change of $T_{N}$ is given as
\begin{eqnarray}
\frac{T_{N}(U,V)}{T_{N}(U,0)}&=\int dV_A\int dV_B J(V_A-V_B)p(V_A)p(V_B)\nonumber\\
&=1+\lambda\Big(\frac{V}{U}\Big)^2
\label{eq:TNeel}
\end{eqnarray}
with a disorder distribution dependent parameter $\lambda$.

The main difference after introducing nonlocal correlations is a reduction of $T_{N}$. I.e., fluctuations beyond the mean
field reduce the transition temperature.  This effect is visible for DF-2nd results and even more pronounced for DF-FLEX.
This comes as no surprise, as Hafermann\cite{Hafermann} found the same behavior for the clean system and,
at least for the clean system, this is in accordance with DCA and QMC\cite{Kent2005} calculations.

The DMFT solution for large $U$ approximately fulfills Eq.~\ref{eq:TNeel}, but the DF solutions deviate. 
We suspect that this is due to the noise in our data.

\subsection{Mott transition in the 3d system}
The following calculations are done for the paramagnetic Hubbard model below $T_{\text{N\'{e}el}}$. This leads to a 
divergence for the FLEX approximation, therefore we have to restrict ourselves to the second-order approximation for the 
dual fermions.

\begin{figure}[tbh]
\centerline{ \includegraphics[clip,scale=0.35]{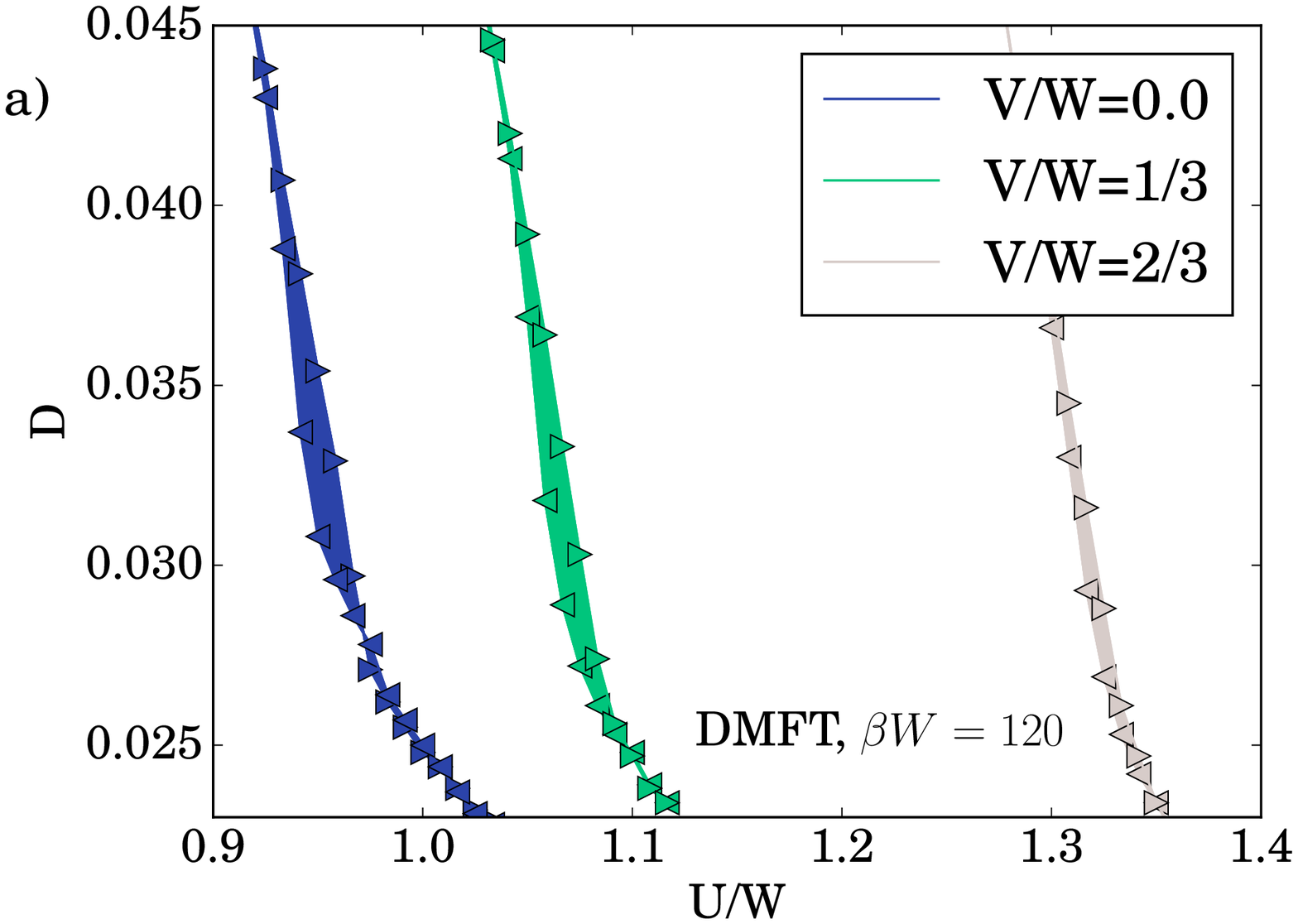}}

\centerline{ \includegraphics[clip,scale=0.35]{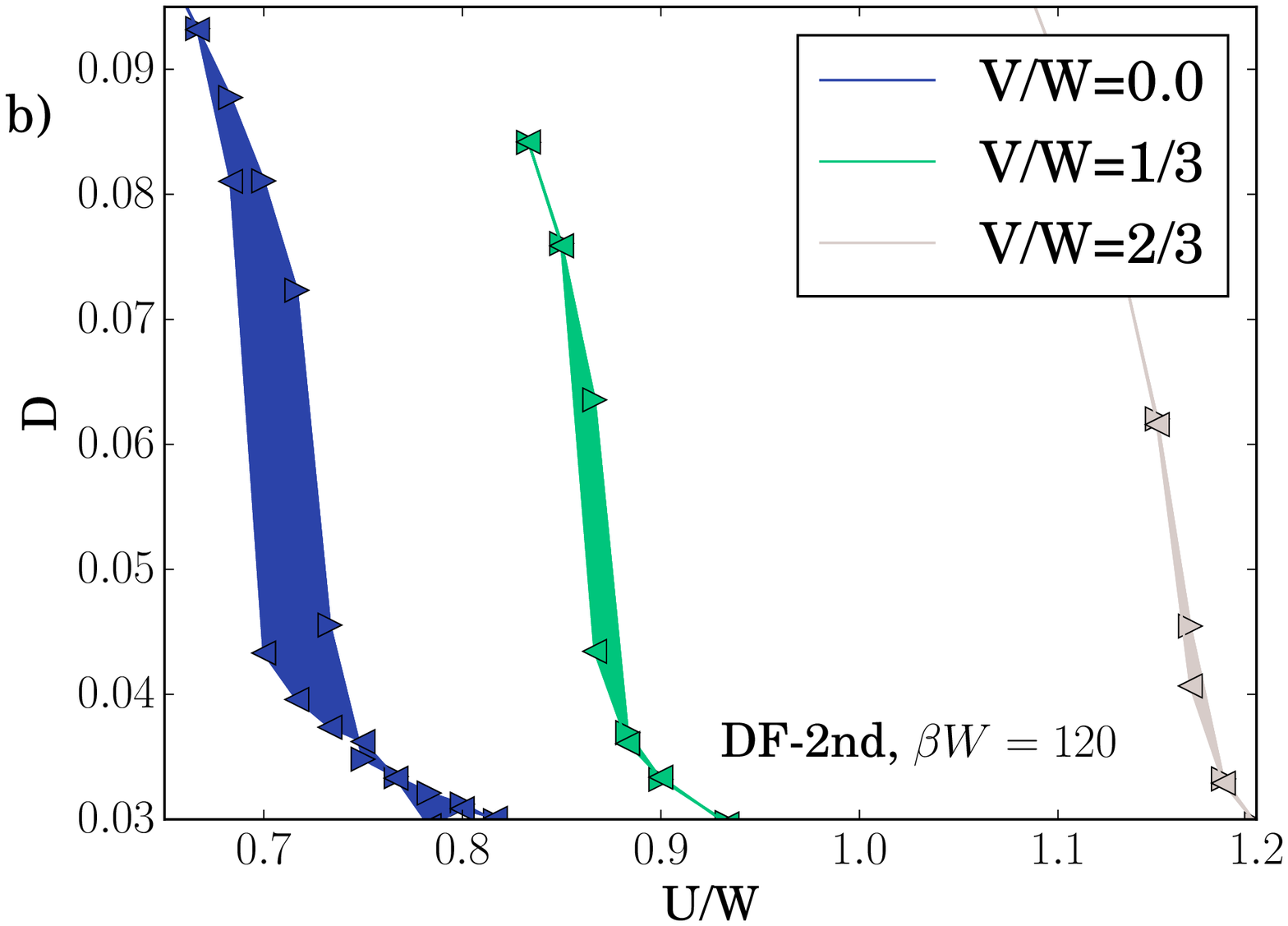}}

\caption{(Color online). The double occupancy $D$ of the impurity problem for DMFT
and 2nd order dual fermions. The double occupancy displays a hysteresis which stems
from the Mott transition. The effect of disorder is to move the hysteresis to larger
values of $U$ and to reduce the area of the hysteresis. The hysteresis from the 
dual fermion calculation is moved to smaller values of $U$ and the area is increased
compared with DMFT+CPA.}
\label{fig:hysterisisB40bin} 
\end{figure}

We investigate the influence of disorder on the Mott transition by looking at the double occupancy $D$ of 
the impurity.  The double occupancy is calculated in the impurity reference system, instead of on the lattice. 
This is due to the missing equation of motion which is present in real fermion systems.  Thus we cannot 
use the trace of single-particle Green function times the self-energy to estimate the double occupancy.
However, the double occupancy measured on the impurity reference system is enough for our purpose
to monitor the hysteresis caused by the first order metal-insulator transition.  We show this
at $\beta W=120$ in Fig.~\ref{fig:hysterisisB40bin} for different values of the disorder strength. 
Fig.~\ref{fig:hysterisisB40bin} shows that disorder moves the hysteresis to larger values of $U$ 
and shrinks the area of the hysteresis. This behavior is captured by DMFT as well, 
but the critical interaction strength is larger for all values of the disorder. 
One can see from the DMFT results that for $V=2/3\,W$ the hysteresis is almost gone,
indicating that strong disorder changes the nature of the Mott transition.
This behavior shows that disorder and interactions compete when it comes to localizing the electrons.

\begin{figure}[tbh]
\centerline{ \includegraphics[clip,scale=0.35]{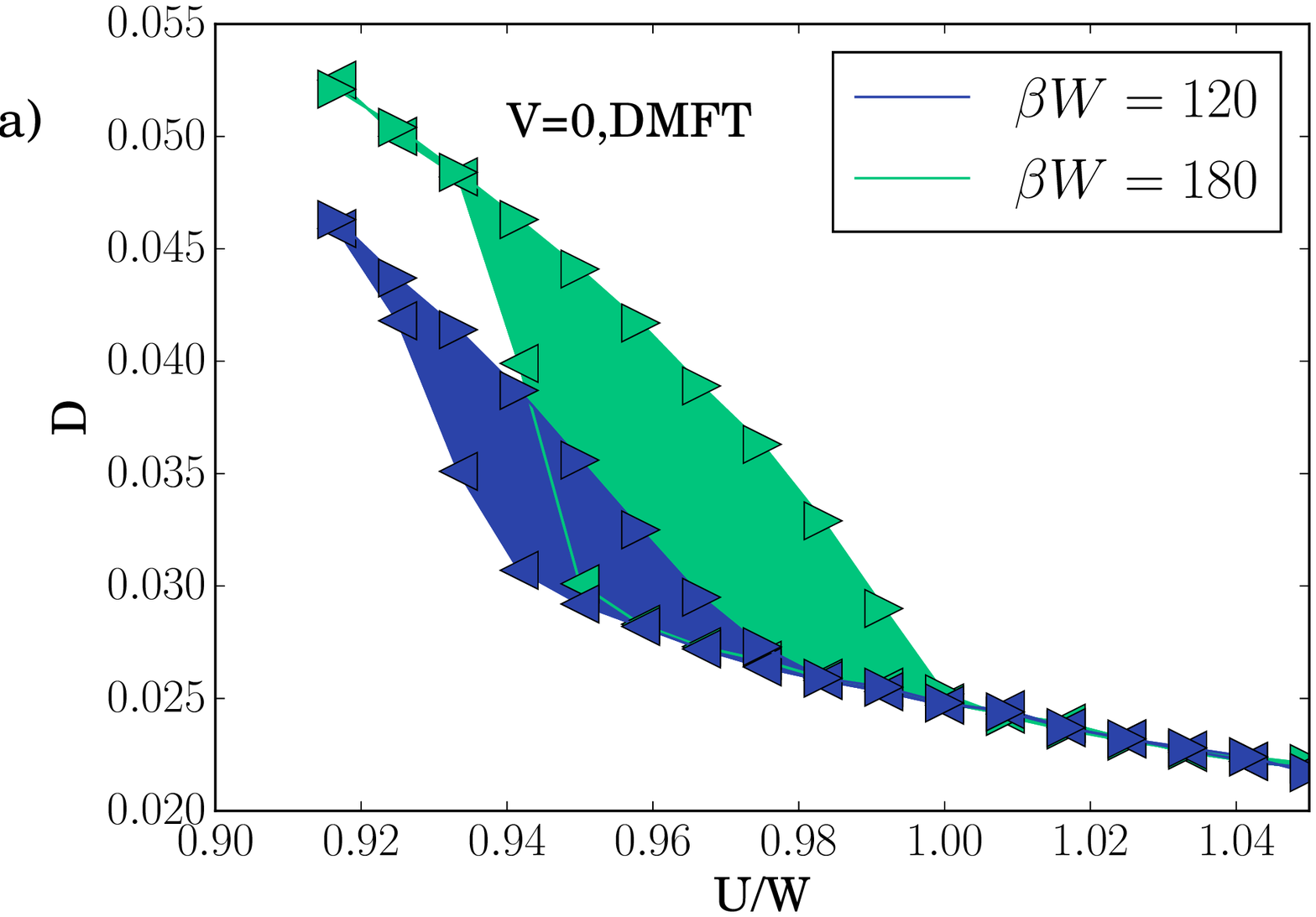}}

\centerline{ \includegraphics[clip,scale=0.35]{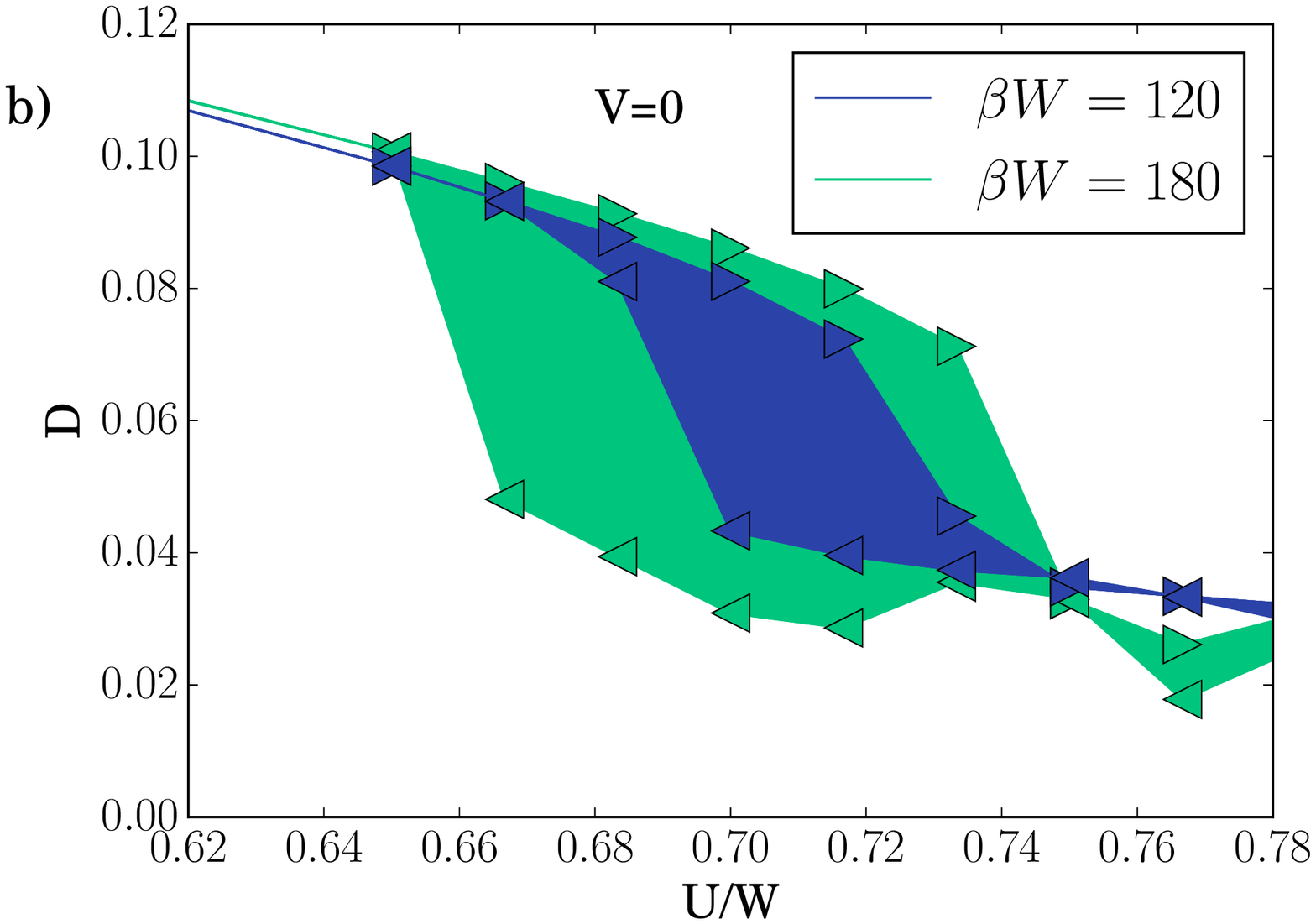}}

\caption{(Color online). Hysteresis at $V=0$ at $\beta W=120$ and $\beta W=180$. 
DMFT predicts that $U_{c_1}$ and $U_{c_2}$ increase with decreasing temperature.
The DF-2nd result shows that $U_{c_1}$ decreases when nonlocal correlations are taken into account.
}
\label{fig:hysteresisV0} 
\end{figure}

\begin{figure}[tbh]
\centerline{ \includegraphics[clip,scale=0.35]{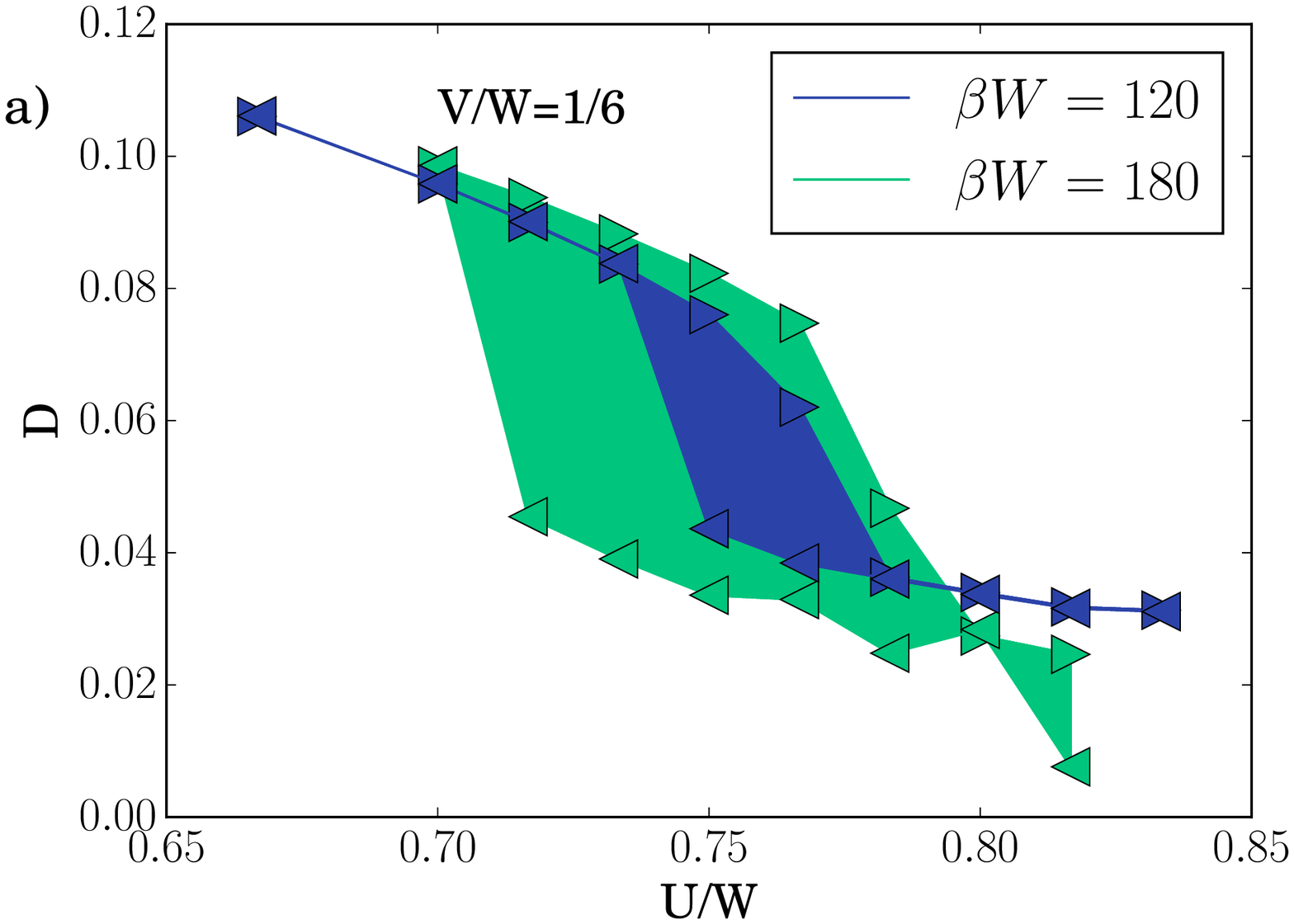}}

\centerline{ \includegraphics[clip,scale=0.35]{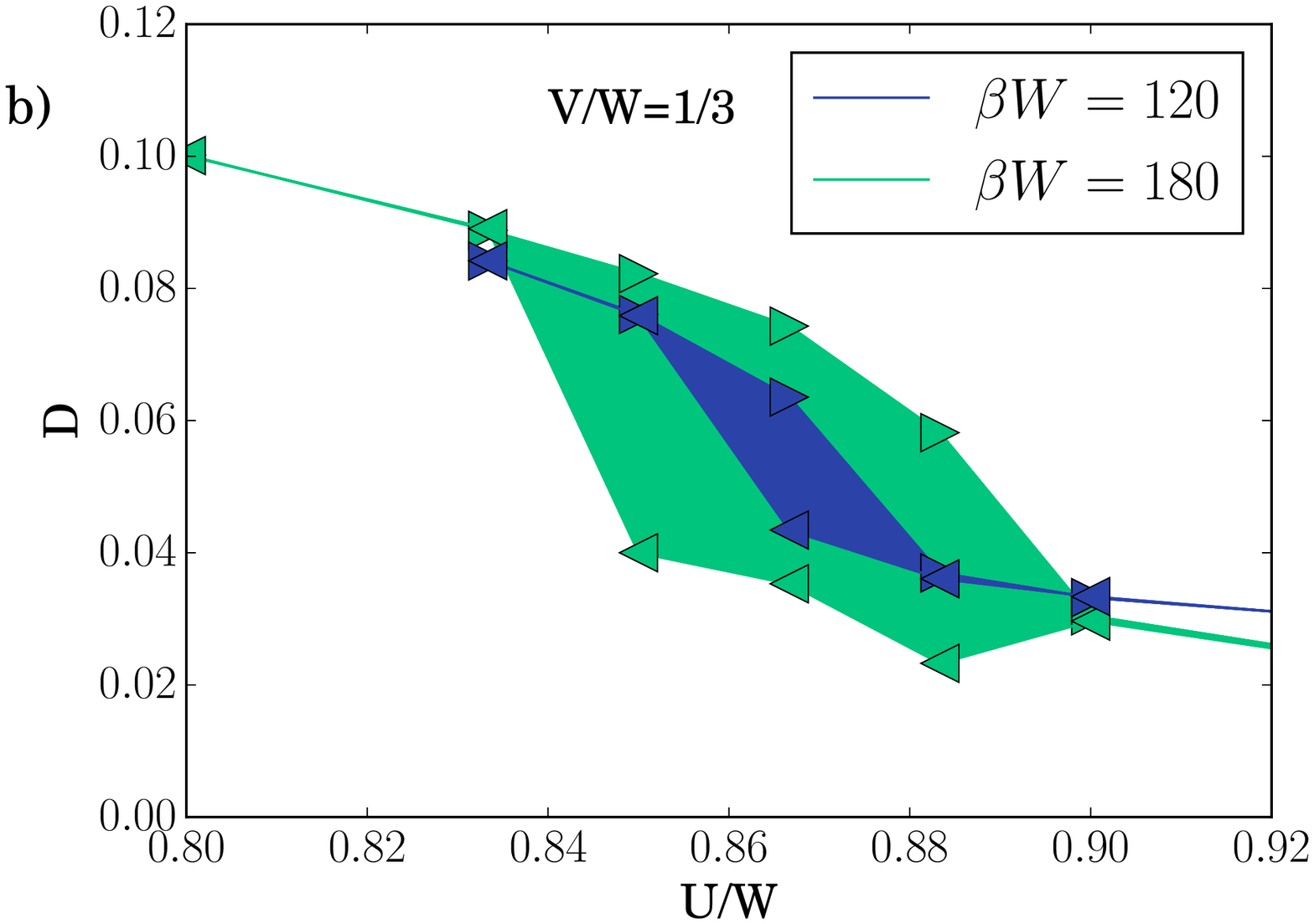}}

\caption{(Color online). DF-2nd results for the hysteresis as a function of temperature at 
$V=\frac{W}{6}$ and $V=\frac{W}{3}$.}
\label{fig:hysterisisVN0} 
\end{figure}

Next, we take a look at the temperature dependence of the hysteresis. In Fig.~\ref{fig:hysteresisV0} 
the hysteresis obtained from DMFT and DF-2nd for the clean system are compared at different temperatures.
DMFT shows mean-field behavior, i.\,e., the upper and lower critical values $U_{c_1}$ and $U_{c_2}$ 
increase with decreasing temperature.  The DF result shows a decreasing $U_{c_1}$ for decreasing temperature. 
Our data is too noisy for the $V=0$ case at large values of $U$ to determine whether $U_{c_2}$ increases or decreases 
with decreasing temperature.
Fig.~\ref{fig:hysterisisVN0} shows DF results for $V=\frac{W}{3}$ and $V=\frac{W}{6}$. For both cases, it is 
clear that $U_{c_2}$ increases with decreasing temperature. $U_{c_1}$ decreases with decreasing temperature, 
just like for the clean system.

\subsection{Phase diagram for the 3d Anderson-Hubbard model}
We calculate the phase diagram on the $UV$ plane for the 3d Anderson-Hubbard model at finite temperature. We explore 
two different quantities. 

\begin{figure}[tbh]
\centerline{ \includegraphics[clip,scale=0.35]{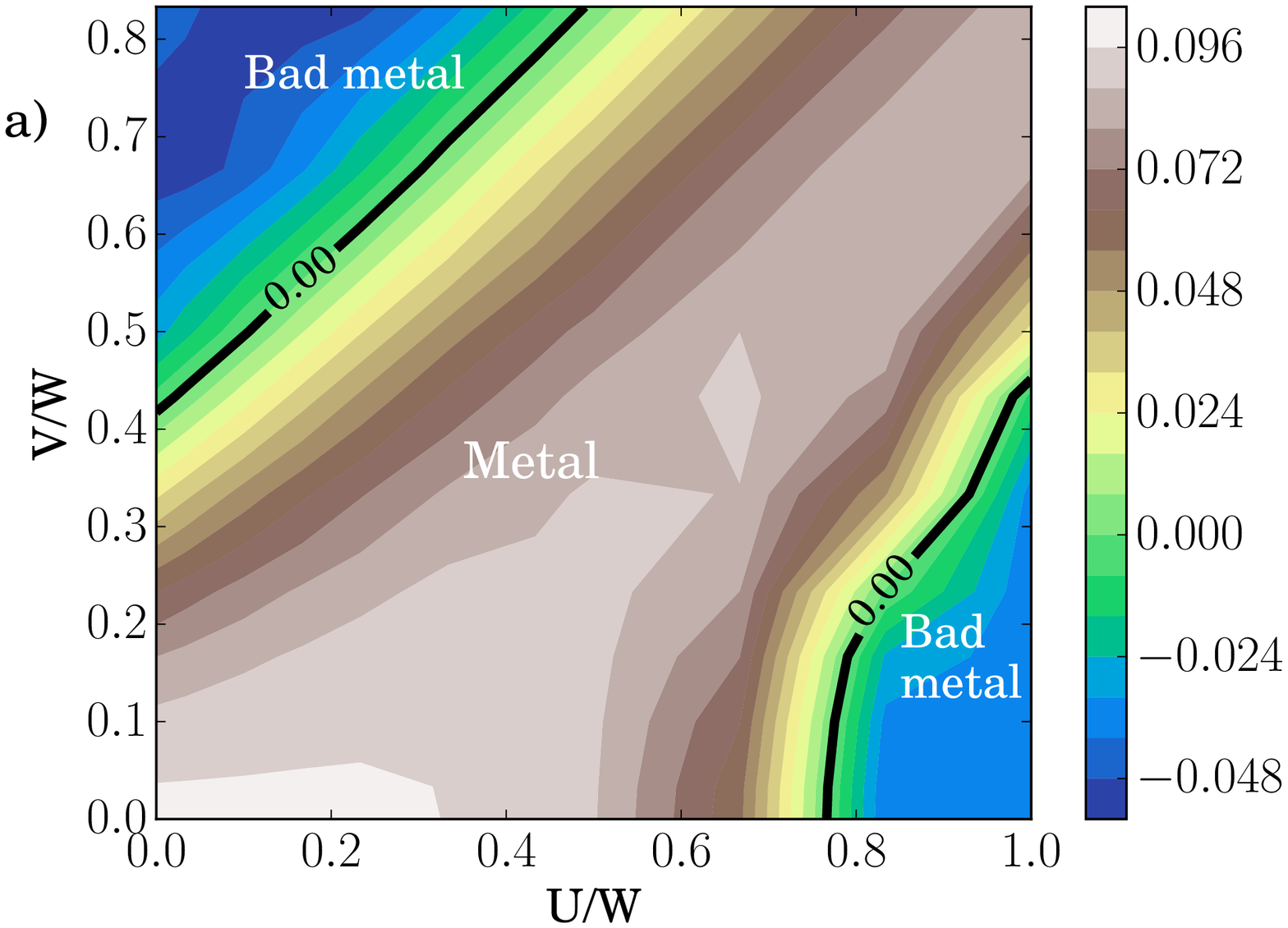}}

\centerline{ \includegraphics[clip,scale=0.35]{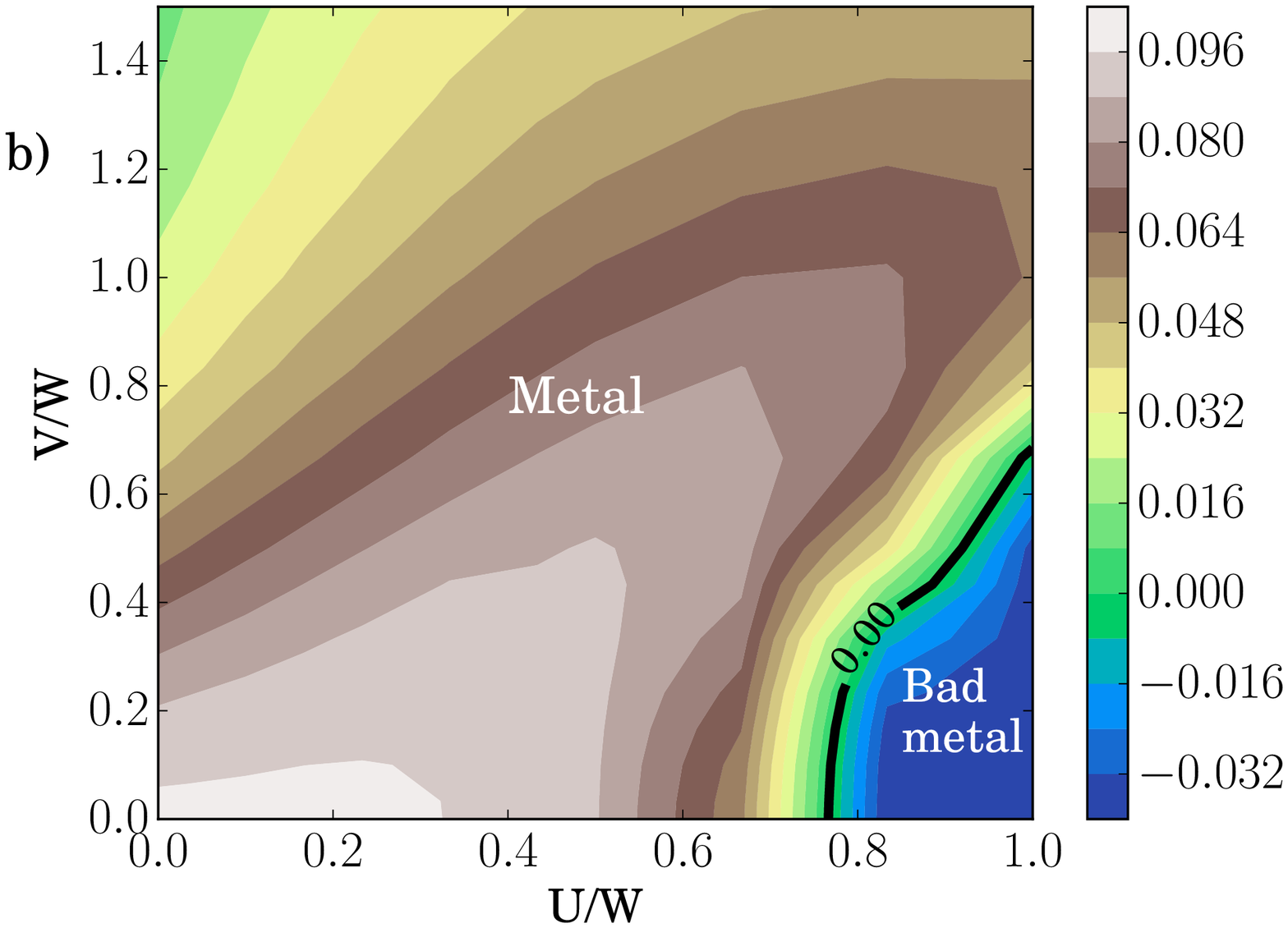}}

\caption{(Color online). $Im\delta G_{loc}$ as a function of $U$ and $V$ at fixed temperature $\beta W=60$ for binary disorder 
(top panel) and box disorder (bottom panel) . We take $Im\delta G_{loc}=0$
as an estimate for the phase boundary. For binary disorder we find an insulating phase for large values of $U$ and for
large values of $V$. For continuous box disorder we still find the insulating phase for large values of $U$ but not
for large values of $V$.}

\label{fig:dg3dB20} 
\end{figure}

The first quantity is the difference $\text{Im}\delta G_{loc}=\text{Im}G_{loc}(3i\pi T)
-\text{Im}G_{loc}(i\pi T)$.  It is only precise for the limit $T\rightarrow 0$, but nevertheless it allows us to detect a 
qualitative difference in the local Green function, namely the presence or lack of a minimum for the imaginary part. 
$\text{Im}\delta G_{loc}=0$ is used as the criterion for
the phase boundary. Fig.~\ref{fig:dg3dB20} shows results for binary and box disorder.  

$\text{Im}\delta G_{loc}$ becomes zero around $U=0.76\,W$. With increasing disorder the Mott transition is moved to 
large values of $U$ for both binary and box disorder.
The details of the phase boundary in this region depend on the disorder distribution but the general behavior is the same.
This picture changes for small $U$ and large $V$. Binary disorder can open a gap and does so around $V_c=0.45\,W$, 
giving rise to an insulating phase for strong disorder.
Box disorder, on the other hand, does not open a gap, which means $\text{Im}\delta G_{loc}<0$ is not possible. Thus, 
we cannot get an estimate for the insulating phase.

\begin{figure}[tbh]
\centerline{ \includegraphics[clip,scale=0.35]{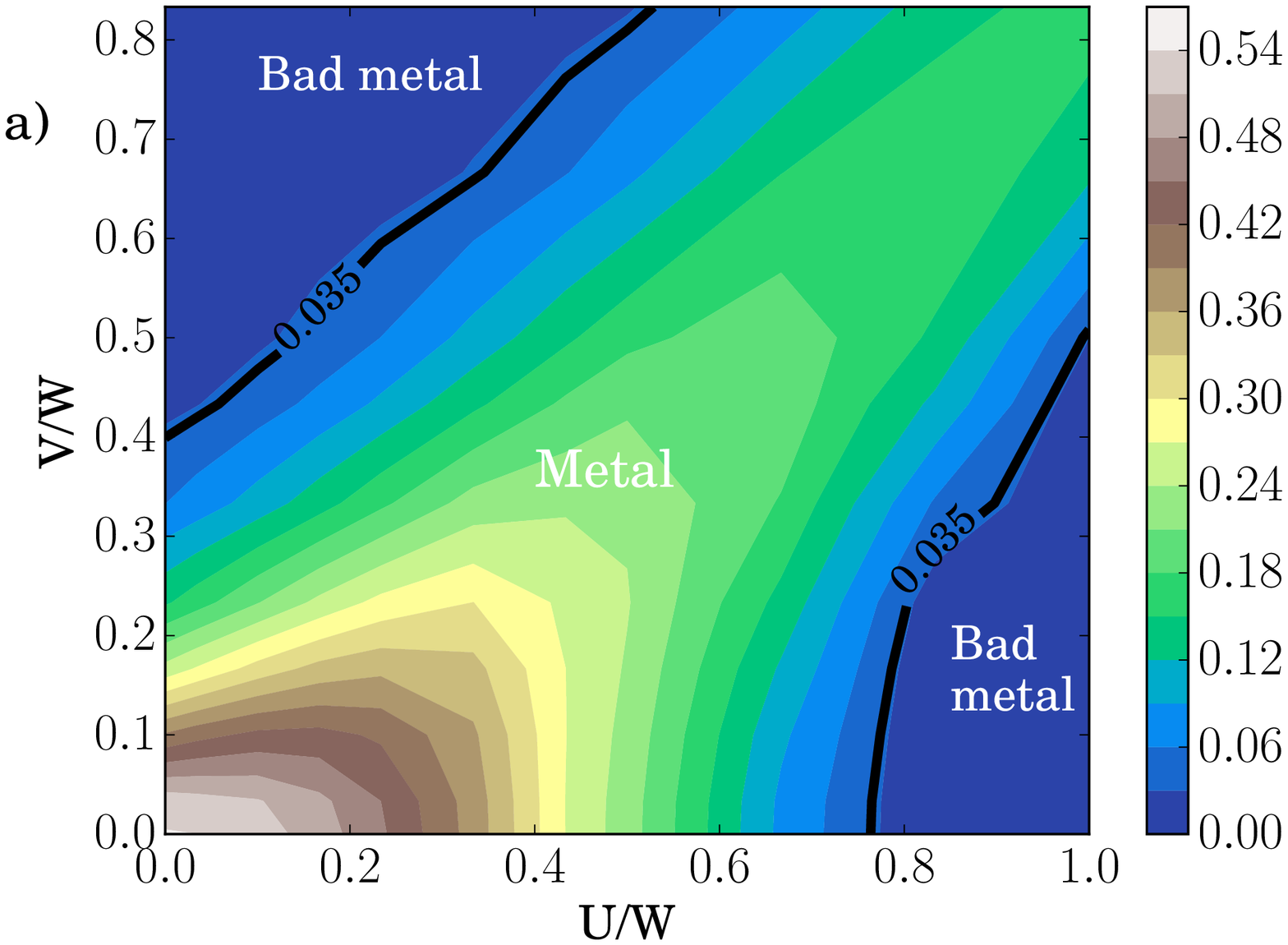}}

\centerline{ \includegraphics[clip,scale=0.35]{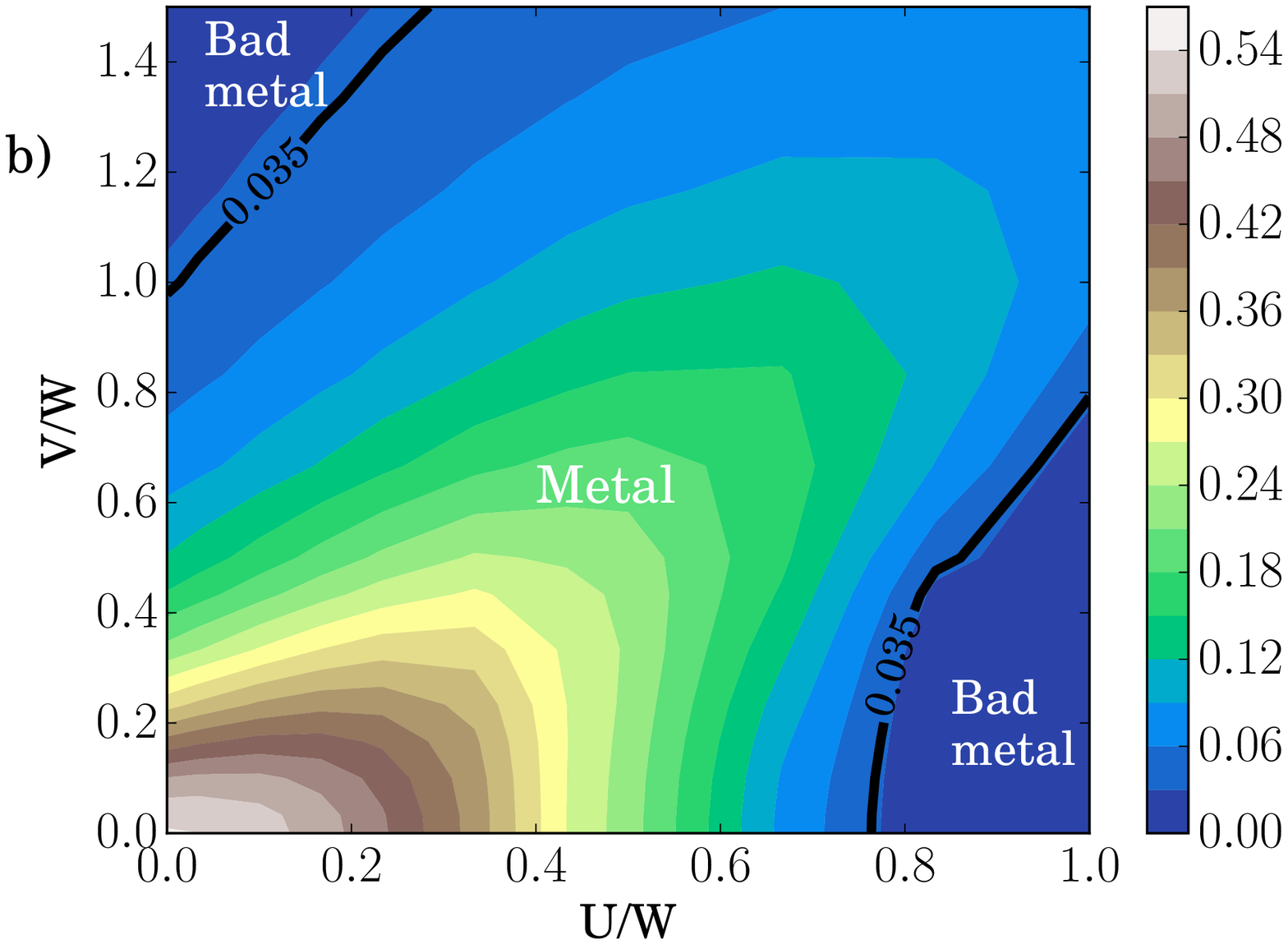}}

\caption{(Color online). Conductivity on the $UV$ plane. For both binary (top panel) and box disorder (bottom panel) 
the conductivity gives a phase transition for large $U$ and large $V$.}

\label{fig:condB20} 
\end{figure}

To overcome this problem the second quantity we explore is the dc conductivity $\sigma_{dc}$, which we calculate 
according to\cite{trivedi96}
\begin{equation}
\sigma_{dc}=\frac{\beta^2}{\pi}\chi_{xx}(q=0,\tau=\frac{\beta}{2}),
\label{eq:sigma_dc}
\end{equation}
where $\chi_{xx}(q,\tau)=\langle j_x(q,\tau)j_x(-q,0)\rangle$ is the current-current correlation function. 
$\chi_{xx}$ is approximated with the bubble diagram and vertex corrections taken into account only involve the pure 
disorder contributions. The conductivity is shown in Fig.~\ref{fig:condB20}. We find that the vertex corrections including interactions become very noisy around the 
transition and we observe a possible lack of thermodynamic consistency.

We use $\sigma_{dc}(U=0.76)=0.035$ for $\beta W=60$ to delineate the boundary of the metallic phase. For both binary and box disorder,
the phase boundary for large $U$ looks similar to the one obtained from $\text{Im}\delta G_{loc}$.
In the case of small $U$ the situation for binary disorder does not change, except for a small reduction of 
the critical disorder strength for $U=0$ to about $0.4\,W$. 
For box disorder, on the other hand, we are now able to determine a phase boundary, which was not possible before,
with $V_c\approx1.0\,W$.
For comparison we want to give the typical medium DCA estimates for $T=0$ by Ekuma et al. \cite{Ekuma2015}.
They found $V_c=0.46\,W$ for binary disorder and $V_c=1.4\,W$ for box disorder. 

We conclude that the DF method at finite temperatures allows one to obtain a reasonable estimate for the Anderson transition, 
but DF with the criteria presented here is not suited to obtain the precise value of $V_c$.


\section{Conclusion}
We discussed the changes needed to apply the dual fermion formalism for disordered interacting systems presented 
in Ref.~\onlinecite{YangHaase} to the Anderson-Hubbard model. The modifications are straight forward, the main difference is
the inclusion of the spin degrees of freedom for the two-particle vertex functions and dual potential.

First, we applied the formalism to the 1d system, which allows for a comparison with DCA calculations for a reasonably 
large cluster size. We found very good agreement with DCA for the relative correction $\sigma(G_{loc})$ to the
local Green function, confirming that DF is indeed able to treat disordered interacting systems and take
into account nonlocal correlations.

Second, we looked at the 3d system. We started with the antiferromagnetic transition.
The phase diagram on the $UT$ plane is in general agreement with the DMFT result on the infinite dimensional
Bethe lattice\cite{Ulmke95}. We compare results from DMFT, DF-2nd and DF-FLEX methods. All three approaches showed 
a suppression of antiferromagnetism for small values of $U$ and disorder. 
For large values of $U$ the approaches agree that weak disorder enhances antiferromagnetism.
The effect of nonlocal correlations from DF was to reduce the transition temperature and the 
reduction was strongest for DF-FLEX. 
The effect of the nonlocal correlations agreed with Hafermann~\cite{Hafermann}.

We continued with the Mott transition. To this end we took a look at the hysteresis of the double occupancy.
Both DMFT and DF show that disorder shifts the transition to larger values of $U$. The effect of
nonlocal correlations was shown to be an overall reduction of the critical $U$.
For the temperature dependence of the hysteresis we found that the DF method gives a qualitatively different result 
than DMFT. DMFT predicts that the lower and upper critical values $U_{c_1}$ and $U_{c_2}$ of the
interaction strength increase with decreasing temperature. The DF result is different in that
it predicts a decreasing $U_{c_1}$ for decreasing temperature. This did not change with the introduction 
of disorder.

Finally, we attempted to calculate a phase diagram on the $UV$ plane. Using the single particle Green function
we were able to get a good guess for the overall shape of the metallic phase, but this method failed
for box disorder. Thus, we calculated the conductivity. The phase diagram for binary disorder remained mostly
unchanged. For box disorder, the conductivity allows to determine the boundary of the metallic phase, which was
not possible from the single particle Green function. For both binary disorder as well as box disorder, 
the critical disorder strength $V_c$ for Anderson localization comes out too small compared to Ref.~\onlinecite{Ekuma2015}.

We conclude that the dual fermion approach for disordered interacting system performs very well, as long as one 
stays away from the disorder-induced metal-insulator transition.

\begin{acknowledgments}
Early parts of this work were supported by DOE SciDAC grant DE-FC02-10ER25916 (SY and MJ) and 
later by NSF EPSCoR Cooperative Agreement No. EPS-1003897 (SY), NSF OISE-0952300 (SY, JM),
and by DFG through research unit FOR 1807 (TP and PH). Computer support is provided by the
Louisiana Optical Network Initiative, by HPC@LSU computing, and by the Gesellschaft 
f\"ur wissenschaftliche Datenverarbeitung G\"ottingen (GWDG) and the GoeGrid project.
\end{acknowledgments}

\appendix
\begin{widetext}
\section{Dual fermion mapping}
\label{sec:df-mapping} 

The derivation of the dual fermion mapping was done for the Anderson-Falicov-Kimball model
previously \cite{YangHaase}. Noting the added complexity of the Anderson-Hubbard model described by 
Eq. \ref{eq:AH} due to  spin indices, in this section, we will re-derive the dual fermion formalism using 
the replica technique.

The disorder averaged lattice Green function is given by 
\begin{equation}
G_\sigma(w,{\bf k}) = -\frac{\delta}{\delta\eta_{w{\bf k}\sigma}}\left\{\ln Z^{v}[\eta_{\omega{\bf k}\sigma}]\right\}|_{\eta_{w{\bf k}\sigma}=0},
\label{GF-defintion}
\end{equation}
 with $\left\{(...)\right\}=\int dvp(v)(...)$ indicating a disorder
averaged quantity, $X^v$ representing the quantity $X$ in disorder configuration $v$ and $\eta_{w{\bf k}\sigma}$ being a source field. 
The partition function for a given disorder configuration $\{v_i\}$ is defined as
\begin{equation}
Z^{v}[\eta_{w{\bf k}\sigma}] = \int D\bar{c}Dce^{-S^{v}[\eta_{w{\bf k}\sigma}]},
\end{equation}
where $Dc\equiv\prod_{w{\bf k}\sigma}{dc_{w{\bf k}\sigma}}$, and the action is itself defined as
\begin{equation}
 S^{v}[\eta_{w{\bf k}\sigma}] = \sum_{w{\bf k}\sigma}\bar{c}_{w{\bf k}\sigma}(-iw+\varepsilon_{{\bf k}}-\mu+\eta_{w{\bf k}\sigma})c_{w{\bf k}\sigma} 
+\sum_{i\sigma}v_{i} \int_0^\beta d\tau n_{i\sigma}(\tau)
 + U\sum_{i} \int_0^\beta d\tau n_{i\uparrow}(\tau)n_{i\downarrow}(\tau),
\label{eq:action}
\end{equation}
 where $iw=i(2n+1)\pi T$ are the Matsubara frequencies, $\varepsilon_{{\bf k}}$
is the lattice bare dispersion, $\mu$ is the chemical potential, and $U$ the Coulomb interaction.
In the following, the explicit functional dependence on source term $\eta_{w{\bf k}\sigma}$ 
for the action will be hidden to simplify the expressions. 
Using the replica trick 
\begin{equation}
\ln Z = \lim_{m\rightarrow0} \frac{Z^{m}-1}{m}, 
\end{equation}
where $m$ replicas are introduced,
we can express the disorder-averaged Green function as 
\begin{equation}
G_\sigma(w,{\bf k})=-\lim_{m\rightarrow0}\frac{1}{m}\frac{\delta}{\delta\eta_{w{\bf k}\sigma}}
\left\{
\int\mathcal{D}\bar{c}\mathcal{D}ce^{-S^{v_i}[c^{\alpha},\bar{c}^{\alpha}]}
\right\}|_{_{\eta_{w{\bf k}\sigma}=0}},
\label{app-eq:GF}
\end{equation}
where $\mathcal{D}c\equiv\prod_{w{\bf k}\sigma\alpha}{dc_{w{\bf k}\sigma}^{\alpha}}$, 
and $\alpha$ is the replica index. The replicated lattice action  is
\begin{equation}
S^{v_i}[c^{\alpha},\bar{c}^{\alpha}]
  = \sum_{w{\bf k}\sigma\alpha}\bar{c}_{w{\bf k}\sigma}^{\alpha}(-iw+\varepsilon_{{\bf k}}-\mu+\eta_{w{\bf k}\sigma})c_{w{\bf k}\sigma}^{\alpha} 
  +  \sum_{i\alpha\sigma}v_{i}\int_{0}^{\beta}d\tau n_{i\sigma}^{\alpha}(\tau)  
  + U\sum_{i\alpha}\int_{0}^{\beta}d\tau n_{i\uparrow}^{\alpha}(\tau) n_{i\downarrow}^{\alpha}(\tau).
\label{app-eq:action_original}
\end{equation}
The disorder averaging can be formally done, and thus we obtain 
\begin{equation}
S[c^{\alpha},\bar{c}^{\alpha}]
  = \sum_{w{\bf k}\sigma\alpha}\bar{c}_{w{\bf k}\sigma}^{\alpha}(-iw+\varepsilon_{{\bf k}}-\mu+\eta_{w{\bf k}\sigma})c_{w{\bf k}\sigma}^{\alpha} 
  +  \sum_{i}W(\tilde{n}_{i}) 
   + U\sum_{i\alpha}\int_{0}^{\beta}d\tau n_{i\uparrow}^{\alpha}(\tau) n_{i\downarrow}^{\alpha}(\tau). 
\label{app-eq:action_av}
\end{equation}
Note that the Coulomb interaction term remains the same, and a new elastic, effective interaction between
electrons of different replicas $W(\tilde{n}_{i})$ appears due to the disorder scattering. 
The latter is local in space and nonlocal in time, and could be expressed through local cumulants $<v_i^{l}>_{c}$ as \cite{m_jarrell_01a}
\begin{equation}
e^{-W(\tilde{n}_{i})} = \int dv_{i}p(v_{i})e^{-v_{i}\sum_{\alpha \sigma}\int d\tau n_{i \sigma}^{\alpha}(\tau)}
  =  e^{-\sum_{l=2}^{\infty}\frac{1}{l!}<v_i^{l}>_{c}\left(\sum_{\alpha \sigma}\int d\tau n_{i \sigma}^{\alpha}(\tau)\right)^{l}}.
\label{eq:cumulant}
\end{equation}

Similarly to the non-interacting disorder fermionic systems~\cite{h_terletska_13}, we follow four steps
to derive the DF formalism for the interacting disorder models.
First, we introduce an effective single-site impurity reference problem
by formally rewriting the original action as 
\begin{equation}
S=\sum_{i}S_{imp}[\bar{c}_i^{\alpha},c_i^{\alpha}]-\sum_{w{\bf k}\sigma\alpha}{\bar{c}_{w{\bf k}\sigma}^{\alpha}(\Delta_{w}-\varepsilon_{{\bf k}}-\eta_{w{\bf k}\sigma})c_{w{\bf k}\sigma}^{\alpha}},
\label{eq:action_with_imp}
\end{equation}
 with an effective impurity action (containing both the Coulomb and disorder interactions,
$W(\tilde{n}_{i}$)) 
\begin{equation}
S_{imp} = \sum_{w\sigma\alpha}\bar{c}_{wi\sigma}^{\alpha}(-iw-\mu+\Delta_{w})c_{wi\sigma}^{\alpha}
 + W(\tilde{n}_{i}) + U\sum_{\alpha}\int_{0}^{\beta}d\tau n_{i\uparrow}^{\alpha}(\tau) n_{i\downarrow}^{\alpha}(\tau).
\end{equation}
 Here $\Delta_{w}$ is a local, and yet unknown, hybridization function
describing the interaction of the impurity with the effective medium.
As in the original DF formalism, it is assumed
that all the properties of the impurity problem, i.\,e., the one-particle
Green function \begin{equation}
G_{imp,\sigma}(w)=-\lim_{m\rightarrow0}\frac{1}{m}\sum_{\alpha=1}^{m}{\displaystyle \int\mathcal{D}\bar{c}\mathcal{D}c\,
 c_{w\sigma}^{\alpha}\bar{c}_{w\sigma}^{\alpha}e^{-S_{imp}},}
\end{equation}
 and the two-particle Green functions which contain effects from both Coulomb
interaction and disorder
\begin{equation}
 \chi_{\sigma_1\sigma_2\sigma_3\sigma_4}^p(\nu)_{w,w'}
= \lim_{m\rightarrow0}\frac{1}{m}
\sum_{\alpha,\beta,\gamma,\delta=1}^{m}{\displaystyle \int\mathcal{D}\bar{c}\mathcal{D}c \,
c_{w+\nu,\sigma_1}^{\alpha}c_{-w,\sigma_2}^{\beta}\bar{c}_{-w',\sigma_4}^{\gamma}\bar{c}_{w'+\nu,\sigma_3}^{\delta}\, e^{-S_{imp}}} 
\end{equation}
 can be calculated. 
These Green functions are local quantities. Our
task is to express the original lattice Green function and other properties
via quantities of the DMFT+CPA impurity problem. What has been accomplished
so far in Eq.~$(\ref{eq:action_with_imp})$ is that the local part
of the lattice action has been moved to the effective impurity.

At the second step of the DF procedure we introduce auxiliary
({}``dual'' fermions) degrees of freedom. In doing so, we transfer
the nonlocal part of the action in Eq.~$(\ref{eq:action_with_imp})$
to the dual variables. As a result, the original real fermions carry
information about the local part only. The transformation to dual
fermions is done via a Gaussian transformation of the nonlocal part
of Eq.~$(\ref{eq:action_with_imp})$,

\begin{equation}
 e^{\bar{c}_{w{\bf k}\sigma}^{\alpha}A_{w{\bf k}\sigma}^{2}c_{w{\bf k}\sigma}^{\alpha}}
=\frac{A_{w{\bf k}\sigma}^{2}}{\lambda_{w\sigma}^{2}}\int\mathcal{D}\bar{f}\mathcal{D}f
e^{-\lambda_{w\sigma}(\bar{c}_{w{\bf k}\sigma}^{\alpha}f_{w{\bf k}\sigma}^{\alpha}+\bar{f}_{w{\bf k}\sigma}^{\alpha}c_{w{\bf k}\sigma}^{\alpha})
-\frac{\lambda_{w}^{2}}{A_{w{\bf k}\sigma}^{2}}\bar{f}_{w{\bf k}\sigma}^{\alpha}f_{w{\bf k}\sigma}^{\alpha}},\label{eq:Hub-Strat}
\end{equation}
 with $A_{w{\bf k}\sigma}^{2}=(\Delta_{w}-\varepsilon_{{\bf k}}-\eta_{w{\bf k}\sigma})$, and $\lambda_{w\sigma}$
yet to be specified.

With such a transformation, the lattice Green function of Eq.~$(\ref{app-eq:GF})$
can be rewritten as 
\begin{eqnarray}
G_\sigma(w,{\bf k}) & = & -\lim_{m\rightarrow0}\frac{1}{m}\frac{\delta}{\delta\eta_{w{\bf k}\sigma}}\frac{\left(\Delta_{w}-\varepsilon_{{\bf k}\sigma}-\eta_{w{\bf k}\sigma}\right)}{\lambda_{w\sigma}^{2}}
\int\mathcal{D}\bar{f}\mathcal{D}f\, e^{-\sum_{w{\bf k}\sigma\alpha}\lambda_{w\sigma}^{2}\bar{f}_{w{\bf k}\sigma}^{\alpha}\left(\Delta_{w}-\varepsilon_{{\bf k}\sigma}-\eta_{w{\bf k}\sigma}\right)^{-1}f_{w{\bf k}\sigma}^{\alpha}}\nonumber \\
 & \times & \int\mathcal{D}\bar{c}\mathcal{D}c\, e^{-\sum_{i}S_{site}^{i}[\bar{c}_{i}^{\alpha},c_{i}^{\alpha};\bar{f}_{i}^{\alpha},f_{i}^{\alpha}]}|_{_{\eta_{w{\bf k}\sigma}=0}},\nonumber \\
\label{GF_with_S_site}\end{eqnarray}
in which the replicated action for site $i$ is of the form \begin{equation}
S_{site}^{i}=S_{imp}+\sum_{\alpha w\sigma}\lambda_{w\sigma}\left(\bar{c}_{iw\sigma}^{\alpha}f_{iw\sigma}^{\alpha}+\bar{f}_{iw\sigma}^{\alpha}c_{iw\sigma}^{\alpha}\right).\label{S_site}\end{equation}
 In Eq.~$(\ref{GF_with_S_site})$ the inter-site hopping is transferred
to a coupling between dual fermions.

At the third step of the DF mapping, we integrate out the real fermions
from the local site action $S_{site}^{i}$  separately for each site $i$,
i.\,e., 
\begin{equation}
\int\prod_{\alpha w\sigma}d\bar{c}_{i\sigma}^{\alpha}dc_{i\sigma}^{\alpha}e^{-S_{site}[\bar{c}_{i\sigma}^{\alpha},c_{i\sigma}^{\alpha};\bar{f}_{i\sigma}^{\alpha},f_{i\sigma}^{\alpha}]}
 = Z_{imp}e^{-\sum_{w\alpha\sigma}\lambda_{w\sigma}^{2}G_{imp,\sigma}(w)\bar{f}_{iw\sigma}^{\alpha}f_{iw\sigma}^{\alpha}-V^{d,i}_{\alpha,\beta}[\bar{f}_{i}^{\alpha},f_{i}^{\alpha};\bar{f}_{i}^{\beta},f_{i}^{\beta}]},\label{Vdf_def}
 \end{equation}
 in which $Z_{imp}$ is the partition function for the replicated
impurity system
\begin{equation}
Z_{imp}=\int\prod_{\alpha w\sigma}d\bar{c}_{i\sigma}^{\alpha}dc_{i\sigma}^{\alpha}e^{-S_{imp}[\bar{c}_{i}^{\alpha},c_{i}^{\alpha}]}.
\end{equation}
As in the clean case,
formally this can be done up to infinite order, which makes the mapping
to the DF variables exact. Choosing for convenience $\lambda_{w}=G_{imp}^{-1}(w)$,
the lowest-order of the replicated DF potential $V^{d,i}_{\alpha,\beta}[\bar{f}_{i}^{\alpha},f_{i}^{\alpha};\bar{f}_{i}^{\beta},f_{i}^{\beta}]$
 reads as 
\begin{equation}
V^{d,i}_{\alpha,\beta}[\bar{f}_{i}^{\alpha},f_{i}^{\alpha};\bar{f}_{i}^{\beta},f_{i}^{\beta}] 
= \frac{1}{2}V^{p,0}_{\alpha,\beta}(w,w')\bar{f}_{iw}^{\alpha}\bar{f}_{iw'}^{\beta}f_{iw'}^{\beta}f_{iw}^{\alpha} 
+ \frac{1}{4}V^{p,1}_\alpha(\nu)_{w,w'}\bar{f}_{i,w+\nu}^{\alpha}\bar{f}_{i,-w}^{\alpha}f_{i,-w'}^{\alpha}f_{i,w'+\nu}^{\alpha}.
\end{equation}
In the derivation of the dual potential of the clean system a term of the form $\langle \bar c\bar c cc\rangle$ appears.
Here, additional sums over replica indices appear and one obtains
\begin{equation}
\sum_{\alpha\beta\gamma\delta}\langle \bar c^\alpha \bar c^\beta c^\gamma c^\delta\rangle_{imp}=
\sum_{\alpha\beta}\langle \bar c^\alpha \bar c^\beta c^\beta c^\alpha\rangle_{imp},
\end{equation}
where we reduce the number of replica indices by using that only terms with duplicated replica indices 
are finite.
$\langle \bar c^\alpha \bar c^\beta c^\beta c^\alpha\rangle_{imp}$ has two distinct contributions,
terms that only contain the effective interaction from disorder (or no interaction) and terms that additionally contain the Hubbard interaction.
This is illustrated in Fig.~\ref{fig:Vcontributions}.
\begin{figure}[tbh]
\centerline{ \includegraphics[clip,scale=0.65]{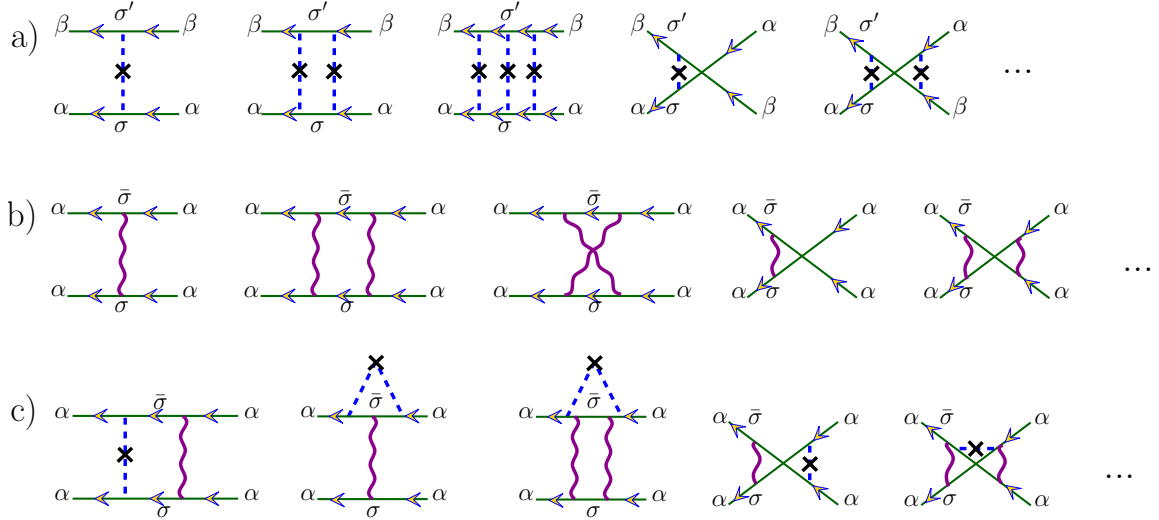}}
\caption{The real fermion impurity diagrams in a) contribute to $V^{p,0}_{\alpha\beta}$, while the diagrams in b) and c) contribute to $V^{p,1}_\alpha$.
For these diagrams all replica labels are fixed to $\alpha$ because of the Hubbard interaction, whereas for diagrams in
a) two replica labels $\alpha$ and $\beta$ remain. $\sigma,\sigma^\prime$ are independent spin labels whereas $\bar\sigma=-\sigma$.}\label{fig:Vcontributions}
\end{figure}
The former interaction acts between different replica, thus these contributions depends on two replica indices
and enter the dual potential in the form of $V^{p,0}_{\alpha,\beta}$.
The latter interaction acts only within one replica, thus these contributions only depend on one replica index.
They enter the dual potential in the form of $V^{p,1}_\alpha$.

In general, the DF vertex $V^{d,i}_{\alpha,\beta}[\bar{f}_{i}^{\alpha},f_{i}^{\beta}]$
contains $n$-body correlation terms introduced by disorder and interaction, but in
the following discussion we will limit ourselves to the leading quartic
term with four external DF fields only.

After taking the derivative with respect to the source field $\eta_{w{\bf k}}$,
the Green function of Eq.~$(\ref{GF_with_S_site})$ reads as 
\begin{equation}
    G_\sigma(w,{\bf k})=\left(\Delta_{w}-\varepsilon_{{\bf k}}\right)^{-1}+\frac{G_{d,\sigma}(w,{\bf k})}{\left(\Delta_{w}-\varepsilon_{{\bf k}}\right)^{2}G_{imp,\sigma}(w)^{2}},\label{origGD}
\end{equation}
 where we define the averaged DF Green function as 
\begin{equation}
    G_{d,\sigma}(w,{\bf k})  =  -\lim_{m\rightarrow0}\frac{1}{m}\sum_{\alpha^{\prime}=1}^{m}\int\mathcal{D}\bar{f}\mathcal{D}f\, e^{-\sum_{w{\bf k}\sigma\alpha}S_{d0}}
   e^{-\sum_{i\alpha\beta w}V^{d,i}_{\alpha,\beta}[\bar{f}_{i\sigma}^{\alpha},f_{i\sigma}^{\alpha};\bar{f}_{i\sigma}^{\beta},f_{i\sigma}^{\beta}]}
   {f}_{w{\bf k}}^{\alpha^{\prime}} \bar f_{w{\bf k}}^{\alpha^{\prime}},\label{GD}
\end{equation}
 and $S_{d0}=\bar{f}_{w{\bf k}\sigma}^{\alpha}\left[{\displaystyle \frac{(\Delta_{w}-\varepsilon_{{\bf k}})^{-1}+G_{imp,\sigma}(w)}{G_{imp,\sigma}^{2}(w)}}\right]f_{w{\bf k}\sigma}^{\alpha}$
is the non-interacting DF action.

Notice, that for the case of non-interacting dual fermions when dual the
potential is zero, Eq.~$(\ref{origGD})$ reduces to the DMFT+CPA solution
for the lattice Green function with $G_\sigma(w,{\bf k})=\frac{1}{G_{imp,\sigma}^{-1}+\Delta_{w}-\varepsilon_{{\bf k}}}$.
Hence, the DMFT+CPA is the zeroth order approximation within our framework.
\end{widetext}

\section{Replica limit}
\label{app:replim}
The replica trick is used to integrate out the disorder in favor of an effective interaction between different replicas.
It is possible to perform the replica limit for the dual fermion diagrams such that the formalism itself does not depend
on replica indices. In this work, the replica trick is used for the purpose of book-keeping, so
that we can derive the dual fermion formalism in a convenient way and non-physical Feynman diagrams can been eliminated 
automatically when taking the replica limit. We would like to emphasize that this does not result in any approximation.

\begin{figure}[tbh]
\centerline{ \includegraphics[clip,scale=0.7]{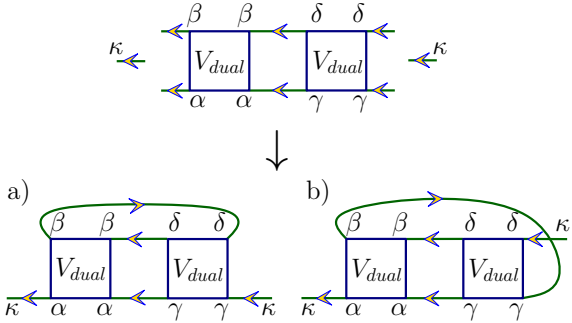}}
\caption{Two possible connections for the second order particle-particle channel diagram for the dual Green function.}\label{fig:repLim2}
\end{figure}
In Fig.~\ref{fig:repLim2} the construction of a second order dual fermion diagram from the vertex ladder is shown.
There are two ways to fix replica indices. First, dual fermions travel only within one replica, i.\,e.
\begin{equation}
\contraction{}{\bar f^\alpha}{}{f^\beta}\bar f^\alpha f^\beta\propto G^\alpha_d\delta_{\alpha\beta}.
\end{equation}
Furthermore, connecting a dual Green function to the potential fixes the involved replica indices, i.\,e.
\begin{equation}
V_{\alpha,\beta}G_d^\gamma=V_{\alpha,\beta}G_d^\alpha\delta_{\alpha\gamma}
\end{equation}
if the Green function line connects to the bottom of the box representing the dual potential or
\begin{equation}
V_{\alpha,\beta}G_d^\gamma=V_{\alpha,\beta}G_d^\beta\delta_{\beta\gamma}
\end{equation}
if the Green function line connects to the top of the box.
This implies that replica indices in diagram b) 
are fixed by the Green function lines alone: $\alpha$ is fixed to $\kappa$ by a Green function line, $\gamma$ is fixed to $\alpha$, $\beta$ to $\gamma$ 
and $\delta$ to $\beta$. Hence, only one free replica index $\kappa$ survives.
Second, the dual potential $V^{p,1}_\alpha$ has only one replica index. Thus, in diagram a) in Fig.~\ref{fig:repLim2}
all replica indices are fixed to the outer replica index $\kappa$ if at least one $V^{p,1}_\alpha$ is used to evaluate the diagram,
e.\,g. if the vertex ladder reads $V^{p,1}_\alpha\bar\chi_0^{pp,\alpha\beta}V^{p,0}_{\gamma\delta}$ we have $\beta=\alpha$ and all the remaining indices are
fixed by Green function lines as described above.
Due to the crossing symmetry of $V^{p,1}_\alpha$, diagrams a) and b) are equivalent if they contain at least one $V^{p,1}_\alpha$. 
In that case, we find it most convenient to use diagram a). 
As one has to sum over $\kappa$ these diagrams are of order $m$.

Two more diagrams remain, a) and b) containing $V^{p,0}_{\alpha\beta}$ only. In combination with the connection in diagram a) the replica indices at the bottom are
fixed to $\kappa$ and one free replica index $\beta$ remains at the top. Thus, the diagram is of order $m^2$. 
Diagram b), as always, is of order $m$ as we saw above.

\begin{figure}[tbh]
\centerline{ \includegraphics[clip,scale=0.75]{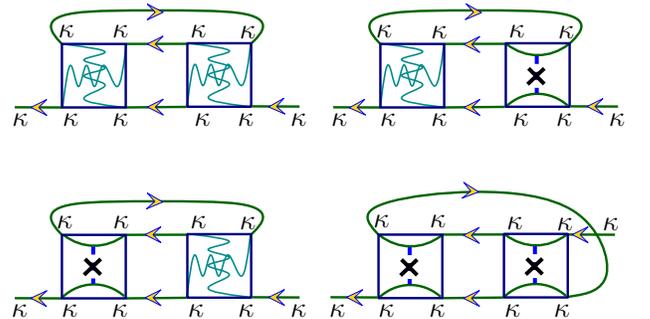}}
\caption{Four diagrams for the second order particle-particle channel are of order $m$, i.\,e. they have only one free replica index $\kappa$.
These four diagrams survive the replica limit $m\rightarrow0$. The crossed wiggly lines represent the crossing-symmetric contributions, whereas the cross with
the curved lines represents the crossing-asymmetric contributions from disorder scattering only.}\label{fig:repLim_result}

\end{figure}

As a result, four diagrams survive the replica limit for the second order contribution in the particle-particle channel. 
These diagrams are shown in Fig.~\ref{fig:repLim_result}.

For the replica limit we have to multiply the diagrams by $\frac{1}{m}$. Thus, diagrams that were of order $m$ are now of order one
and survive the replica limit $m\rightarrow0$. Diagrams that were of order $m^2$ or higher do not survive the replica limit $m\rightarrow0$.
As a result, after the replica limit only the four diagrams displayed in Fig.~\ref{fig:repLim_result} remain for the second order, three 
of type a) and one of type b).

\begin{figure}[tbh]
\centerline{ \includegraphics[clip,scale=0.6]{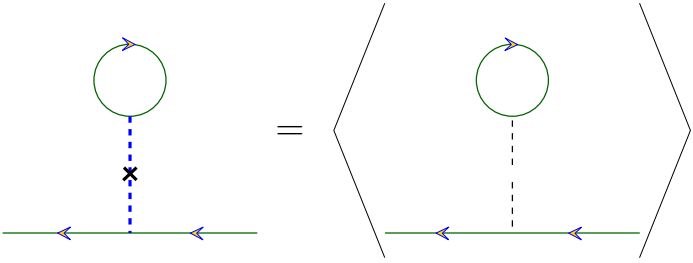}}
\caption{Hartree-like diagrams are created from disorder by disorder averaging disconnected diagrams. 
For quenched disorder all disconnected diagrams are removed before the disorder average, hence such diagrams do not
exist. This property of the real fermion diagrams translates to the dual degrees of freedom.
The black dashed line denotes elastic scattering from an impurity.}\label{fig:dis_hartree}
\end{figure}

With the rules given above, the replica limit can be readily applied to higher order diagrams.
The removal of Hartree-like diagrams can be understood by considering topologically equivalent diagrams 
for the real degrees of freedom. Fig. \ref{fig:dis_hartree} shows the first order Hartree diagram and its 
creation from a disconnected diagram. For quenched disorder, all unconnected diagrams are removed by the
factor $\frac{1}{Z}$ before the disorder average, hence such a diagram does not appear.

\section{Definition of vertex functions}
\label{app:vertexFunctions}
In the main text $V^{ph(pp)}_{\sigma_1\sigma_2;\sigma_3\sigma_4}$ was introduced.
We need the impurity Green function $g_\sigma(\omega)$ for a single disorder configuration and the disorder averaged impurity Green function $G_\sigma(\omega)$
to calculate it.
We obtain
\begin{eqnarray}
V^{ph}_{\sigma_1\sigma_2;\sigma_3\sigma_4}(\nu)_{\omega,\omega^\prime}=&\nonumber\\
&\frac{1}{T}\Big(
\frac{\{ \langle c_{\omega+\nu,\sigma_1}\bar c_{\omega,\sigma_2} c_{\omega^\prime,\sigma_3}\bar c_{\omega^\prime+\nu,\sigma_4}\rangle_{\text{imp}} \}}
{G_{\sigma_1}(\omega+\nu)G_{\sigma_2}(\omega) G_{\sigma_3}(\omega^\prime)G_{\sigma_4}(\omega^\prime+\nu)}\nonumber\\
+&\frac{ G_{\sigma_3}(\omega)G_{\sigma_1}(\omega^\prime+\nu) }
{G_{\sigma_1}(\omega+\nu)G_{\sigma_2}(\omega) G_{\sigma_3}(\omega^\prime)G_{\sigma_4}(\omega^\prime+\nu)}\times\nonumber\\
\times&(\delta_{\sigma_1\sigma_4}\delta_{\sigma_2\sigma_3}\delta_{\omega,\omega^\prime}-\delta_{\sigma_1\sigma_2}\delta_{\sigma_3\sigma_4}\delta_{\nu,0})
\Big)
\nonumber\\
\label{eq:v_ph_appendix}
\end{eqnarray}
for the particle-hole channel and 
\begin{eqnarray}
V^{pp}_{\sigma_1\sigma_2;\sigma_3\sigma_4}(\nu)_{\omega,\omega^\prime}=&\nonumber\\
&\frac{1}{T}\Big(
\frac{\{ \langle c_{\omega+\nu,\sigma_1}c_{-\omega,\sigma_2} \bar c_{-\omega^\prime,\sigma_3}\bar c_{\omega^\prime+\nu,\sigma_4}\rangle_{\text{imp}} \}}
{G_{\sigma_1}(\omega+\nu)G_{\sigma_2}(-\omega) G_{\sigma_3}(-\omega^\prime)G_{\sigma_4}(\omega^\prime+\nu)}\nonumber\\
+&\frac{ G_{\sigma_1}(\omega)G_{\sigma_2}(\omega^\prime+\nu)  }
{G_{\sigma_1}(\omega+\nu)G_{\sigma_2}(-\omega) G_{\sigma_3}(-\omega^\prime)G_{\sigma_4}(\omega^\prime+\nu)}\times\nonumber\\
\times&(\delta_{\sigma_1\sigma_3}\delta_{\sigma_2\sigma_4}\delta_{\omega+\omega^\prime+\nu,0}-\delta_{\sigma_1\sigma_4}\delta_{\sigma_2\sigma_3}\delta_{\omega,\omega^\prime}
\Big)
\nonumber\\
\label{eq:v_pp_appendix}
\end{eqnarray}
for the particle-particle channel.
For convenience we choose a form of $V^{ph(pp)}$ that contains both crossing-symmetric as well as crossing-asymmetric contributions. 
It is possible to remove all crossing-asymmetric contributions from $V^{ph(pp)}$. 
As a consequence, the equations for the dual self-energy in sec. \ref{sec:dfSelfEnergy} would be modified.

\section{Second order dual self-energy}
\label{app:dualSelfEnergy}
For the particle-hole channel there are three possible spin configurations for the second order diagram.
These diagrams are shown in Fig. \ref{fig:2ndOrderPH}. The first diagram contains two equivalent Green function lines,
thus a factor $\frac{1}{2}$ is associated with it. The second and third diagram are topologically equivalent. 
As we want to include both we have to multiply both diagrams with a factor $\frac{1}{2}$ as well.

We want to express the self-energy in terms of $V_{d/m^0}$, thus we use the following relations:
\begin{eqnarray}
V_{\uparrow\uparrow;\uparrow\uparrow}=\frac{1}{2}(V_d+V_m)\\
V_{\uparrow\uparrow;\downarrow\downarrow}=\frac{1}{2}(V_d-V_m)\\
V_{\uparrow\downarrow;\uparrow\downarrow}=V_{m^0}
\end{eqnarray}
The last equality is true because $V_{\uparrow\downarrow;\uparrow\downarrow}$ is part of the triplet channel.

Combining all this together we obtain
\begin{equation}
\begin{split}
\Phi^*=&\frac{1}{2}\Big(\frac{1}{2}(V_d+V_m)\bar\chi_0^{ph}\frac{1}{2}(V_d+V_m)+V_m\bar\chi_0^{ph}V_m \\
&+\frac{1}{2}(V_d-V_m)\bar\chi_0^{ph}\frac{1}{2}(V_d-V_m)\Big)\\
=& \frac{1}{4}(V_d\bar\chi_0^{ph}V_d+3V_m\bar\chi_0^{ph}V_m).
\end{split}
\end{equation}
$\Phi^*$ contains unphysical contributions from the purely disordered contributions. 
To remove all purely disordered contributions we replace $V_{d/m}$ in the above by
their purely disordered counterparts $V_{d/m}^0$ which are defined in Eq. \ref{eq:Vd0} and \ref{eq:Vm0}
and subtract the result from $\Phi^*$. We obtain
\begin{equation}
\Phi=\frac{1}{4}[V_{d}\bar\chi^{ph}_{0}V_{d}+3V_{m}\bar\chi^{ph}_{0}V_{m}]-\frac{1}{4}[V_{d}^{0}\bar\chi^{ph}_{0}V_{d}^{0}+3V_{m}^{0}\bar\chi^{ph}_{0}V_{m}^{0}].
\end{equation}

\begin{figure}[tbh]
\centerline{ \includegraphics[clip,scale=0.8]{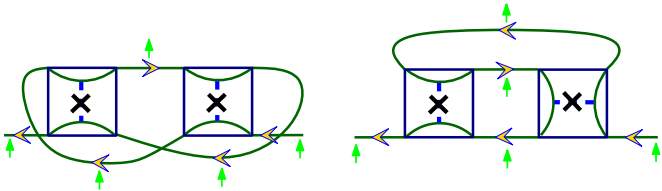}}
\caption{(Color online). Physical 2nd order diagrams for the purely disordered contributions to the particle-hole channel.
Both diagrams are equivalent. The diagram on the left shows that complicated connections are necessary to create
skeleton diagrams for the particle-hole channel. On the right, the artificially introduced vertical disorder vertex is used. 
It is more convenient as it allows to restrict oneself to Hartree-like diagrams. 
This is particularly helpful for higher-order diagrams.}
\label{fig:dis_diag_self_ph} 
\end{figure}

Finally, we have to determine $\Phi^0$. The corresponding diagram is shown in Fig. \ref{fig:dis_diag_self_ph}.
Note that there is only one spin configuration as there is only one dual particle that cannot change its spin.
The result is
\begin{equation}
\Phi^{0}(w,w;q)=\gamma^=(w,w)\bar{\chi}_{0}^{ph}(\nu=0;q)_\omega\gamma^=(w,w).
\end{equation}

Similarly, the self-energy for the particle-particle channel can be calculated, as well as for general 
higher order diagrams. Note that that the symmetry factors required here for the particle-hole channel are
an idiosyncrasy of the second order diagrams and do not appear in higher order ladder diagrams. 
For the particle-particle channel these factors appear at all orders for ladder diagrams.

\bibliography{df_ah}

\end{document}